\documentclass{ws-rmp}

\renewcommand{\Im}{\mathrm{Im\,}}
\renewcommand{\Re}{\mathrm{Re\,}}
\newcommand{\D}{\mathrm{d}}
\newcommand{\e}{\mathrm{e}}
\renewcommand{\i}{\mathrm{i}}
\newcommand{\C}{\mathbb{C}}
\newcommand{\N}{\mathbb{N}}
\newcommand{\R}{\mathbb{R}}
\newcommand{\GG}{\mathcal{G}}
\newcommand{\HH}{\mathcal{H}}
\newcommand{\JJ}{\mathcal{J}}
\newcommand{\OO}{\mathcal{O}}
\newcommand{\RR}{\mathcal{R}}

\begin{document}

\markboth{P. Exner and O. Turek} {Approximations of vertex
couplings in quantum graphs}

%
\catchline{}{}{}{}{}
%

\title{Approximations of singular vertex couplings in quantum graphs}

\author{PAVEL EXNER}

\address{Doppler Institute for Mathematical Physics and Applied
Mathematics, Czech Technical University, B\v rehov{\'a} 7, 11519
Prague, Czech Republic \\
exner@ujf.cas.cz }

\author{OND\v{R}EJ TUREK}

\address{Department of Mathematics, Faculty of Nuclear Sciences and
Physical Engineering, Czech Technical University, Trojanova 13,
12000 Prague, Czech Republic\\
oturek@centrum.cz }

\maketitle

\begin{history}
\received{(Day Month Year)}
\revised{(Day Month Year)}
\end{history}

\begin{abstract}
We discuss approximations of the vertex coupling on a star-shaped
quantum graph of $n$ edges in the singular case when the wave
functions are not continuous at the vertex and no edge-permutation
symmetry is present. It is shown that the Cheon-Shigehara
technique using $\delta$ interactions with nonlinearly scaled
couplings yields a $2n$-parameter family of boundary conditions in
the sense of norm resolvent topology. Moreover, using graphs with
additional edges one can approximate the ${n+1\choose
2}$-parameter family of all time-reversal invariant couplings.
\end{abstract}



\section{Introduction}

The concept of quantum mechanics on a graph is more than half a
century old having roots in modelling of aromatic hydrocarbons
\cite{RS53}. For many years, however, it was rather a curiosity,
or maybe an interesting textbook example. The situation changed
two decades ago with the advent of microfabrication techniques
which allow us to produce tiny graph-shaped structures of
semiconductor and other materials for which this is a useful and
versatile model. This motivated a new theoretical attention to the
subject -- see, e.g., \cite{GP88, ES89}. Since then the literature
on quantum graphs grew to a formidable volume, and we restrict
ourselves here to mentioning recent reviews in \cite{AGHH, Ku04,
BCFK} where an extensive bibliography can be found.

From the mathematical point of view the attractive feature of the
model is that it deals with families of ordinary differential
equations, the solutions of which have to be properly matched at
the graph edge endpoints. Since the solutions are often explicitly
known, the spectral analysis can be reduced to an algebraic
problem.

The key point here are the boundary conditions through which the
wave functions are matched. The Hamiltonian is typically a
second-order differential operator, for instance, in the simplest
case of a free spinless particle it acts on the $j$-th edge as
$H\psi_j= -\psi''_j$. Thus the boundary conditions are linear
relations coupling the values of the functions and their first
derivatives at graph vertices; from the physical point of view it
is usually sufficient to consider only \emph{local} couplings
which involve values at a single vertex only. Another general
physical restriction is the self-adjointness of the Hamiltonian;
it implies that a vertex joining $n$ graph edges may be
characterized by boundary conditions involving $n^2$ real
parameters \cite{ES89}.

This leaves a considerable freedom in the choice of a model to
describe particular physical systems, and an understanding of the
physical meaning of vertex coupling is needed to pick the
appropriate operator from the class of admissible Hamiltonians. A
natural way to approach this problem is through approximation,
i.e. regarding the quantum graph in question as a limit of a
family of more ``realistic'' systems with a less number of free
parameters. One possibility is to approximate a graph by a family
of ``fat graphs'' or similar manifolds equipped with the
corresponding Laplace-Beltrami operators. The best studied case is
the one where the approximated manifolds have Neumann boundary, or
no boundary et all \cite{FW93, KZ01, RS01, Sa01, EP05, Po05a},
where unfortunately the limit yields -- of the multitude of
available boundary conditions -- only the most simple ones. There
are also fresh results \cite{Po05b, MV06} on the case with
Dirichlet boundary but in general the approach based on squeezed
manifolds did not yield so far a satisfactory answer to the
question.

Another, maybe less ambitious approach is to model vertex boundary
conditions through families of interactions on the graph itself.
Here two cases have to be distinguished. In the $n^2$-parameter
family mentioned above the boundary conditions with wave functions
\emph{continuous at the vertex} form just one-parameter subfamily.
These boundary conditions can be approximated by families of
scaled potentials in analogy is analogy with one-dimensional
$\delta$ interactions \cite{E96}. The remaining, \emph{more
singular} cases require a different approach. An inspiration may
be derived from the approximation of one-dimensional $\delta'$
interactions suggested, somewhat surprisingly, by Cheon and
Shigehara in \cite{CS98} and elaborated in a mathematically
consistent way in \cite{AN00, ENZ01}. It is based on a family of
$\delta$ interactions which approach each other being scaled in a
particular nonlinear way. An analogous procedure for vertices of
degree $n\ge 2$ was proposed in \cite{CE04} in the case of the
so-called $\delta'_s$ coupling; the key element here was the
symmetry with respect to permutation of the edges which allowed to
reduce the analysis to a one-dimensional halfline problem. The
same technique was afterwards in \cite{ET06} applied to the class
of all permutation-symmetric boundary conditions which form a
two-parameter subfamily in the $n^2$-parameter set.

The main goal of the present paper is to explore whether the idea
of \cite{CS98} can be adapted to situations without a permutation
symmetry and how wide class of boundary conditions can be in this
way described. As in the work mentioned above we will consider a
\emph{star graph} with a single vertex and $n$ semi-infinite
edges. For simplicity we will also assume that the motion on
graphs edges is \emph{free}; the obtained approximations extend
easily to Schr\"odinger operators on the graph provided the
potentials involved are sufficiently regular around the vertex. We
are going to show that the Cheon-Shigehara technique can produce
for $n>2$ at most a $2n$-parameter family of boundary conditions
at the vertex. Furthermore, we will demonstrate that such an
approximations, with two $\delta$ interaction at each edge, do
indeed exist and that they converge in the norm resolvent
topology.

The next question is how to extend the approximation to a wider
class of couplings. A natural possibility is amend the star by
extra edges supporting $\delta$ interactions which shrink to the
``main'' vertex with the parameter controlling the approximation.
We devise such a scheme a show that it yields an ${n+1\choose
2}$-parameter family, generically \emph{all} couplings which are
\emph{time-reversal invariant}. In this case, however, we restrict
ourselves to deriving the boundary condition formally. We are
convinced that the norm resolvent convergence could be verified as
in the case mentioned but the argument would be extremely
cumbersome. Notice that the idea of using additional edges to
model singular couplings appeared already in \cite{AEL94}. In
contrast to that paper, however, we keep here the number of added
edges fixed.

Let us review briefly the contents of the paper. In the next
section we gather the needed preliminary information. We review
the quantum graph concept, recall different vertex couplings and
review briefly the known approximations. In Section~3 we analyze a
CS-type approximation to the vertex in a star graph based on
adding $\delta$ interactions on star edges, the following section
is devoted to the proof of norm-resolvent convergence. Finally, in
Section~5 we will describe the mentioned more general
approximation with extra edges added to the star graph.

\section{Preliminaries}

\subsection{Quantum graphs}

Let us first recall a few basic notions. A \emph{graph} $\Gamma$
is an ordered pair $\Gamma=(V,E)$, where $V$ and $E$ are finite or
countably infinite sets of \emph{vertices} and \emph{edges},
respectively. Without loss of generality we may identify $E$ with
a family of two-element subsets in $V$, excluding thus loops and
multiple edges, since in the opposite case we can simply add extra
vertices. The vertex \emph{degree} of $v\in V$ is the number of
edges which have $v$ as its endpoint. $\Gamma$ is a \emph{metric
graph} if each of its edges can be equipped with a distance, i.e.
identified with a finite or semi-infinite interval of length
$\ell\in(0,+\infty]$; the endpoints ``at infinity'' are
conventionally not counted as vertices. In particular a \emph{star
graph} has a finite number $n\ge 2$ of edges and a single
\emph{centre} which is the only vertex where all the edges (called
also \emph{arms} in this case) meet.

The subject of our interest is quantum mechanics on graphs. Given
a metric graph  $\Gamma$ with edges $J_1,\dots,J_n$ we identify
the orthogonal sum $\HH= \bigoplus_{j=1}^n L^2(J_j)$ with the
state Hilbert space, i.e. the wave function of a spinless particle
``living'' on $\Gamma$ can be written as the column $\Psi=(\psi_1,
\psi_2,\ldots,\psi_n)^T$ with $\psi_j\in L^2(J_j)$. In the absence
of external fields the Hamiltonian $H$ acts as $(H_\Gamma \Psi)_j=
-\psi''_j$, where as usual we put $\hbar=2m=1$. Its domain
consists of functions from $W^{2,2}(\Gamma) := \bigoplus_{j=1}^n
W^{2,2}(J_j)$; since $H$ is required to be a self-adjoint operator
they must satisfy appropriate boundary conditions at the vertices
which we will recall below.

The meaning of these boundary condition is our main concern in
this paper, therefore we restrict ourselves to graphs with a
single vertex, namely star graphs with $n$ semi-infinite edges
$J_j\simeq\R^+,\: j=1,\dots,n$; we denote them as $\Gamma$ or
$\Gamma_n$.

\subsection{Vertex couplings}

Since the Hamiltonian mentioned above is a second-order operator,
the matching conditions involve boundary values of the functions
in the vertex and of their first derivatives. Both regarded as
one-sided limits, the derivatives are taken in the outward
direction. We arrange them into column vectors $\Psi(0)$ and
$\Psi'(0)$. The self-adjointness of $H$, which in the physical
language means conservation of probability current at the vertex,
is expressed through a linear relation between these vectors,
\begin{equation}\label{1}
A\Psi(0)+B\Psi'(0)=0\,,
\end{equation}
by \cite{KS99} the operator $H$ is self-adjoint if and only if
$A,B\in\C^{n,n}$ satisfy the conditions
\begin{equation}\label{OP}
\mathrm{rank}(A,B)=n\,, \quad AB^* \textrm{\;\;is self-adjoint},
\end{equation}
where $(A,B)$ denotes the $n\times2n$ matrix with $A,B$ forming
the first and the second $n$ columns, respectively. This
parametrization is obviously non-unique, since $A,B$ can be
replaced by $CA,CB$ with any regular $n\times n$ matrix $C$. This
defect can be corrected by choosing the matrices in the standard
form \cite{Ha00, KS00},
 \begin{equation}\label{parametrizace}
 (U-I)\Psi(0)+\i(U+I)\Psi'(0)=0\,,
 \end{equation}
where $U$ is an $n\times n$ unitary matrix; the Hamiltonian
corresponding to this condition will be labelled as $H_U$.
Elements of this family are labelled by $n^2$ real parameters
which is, of course, the right number because all the $H_U$ are
self-adjoint extensions of a common symmetric restriction with
deficiency indices $(n,n)\,$ \cite{ES89}.

Let us next recall a few examples of the boundary conditions
(\ref{parametrizace}). As mentioned in the introduction, the
requirement of continuity at the vertex selects a one-parameter
subfamily corresponding to the so-called \emph{$\delta$ coupling},
\begin{equation}\label{delta}
\psi_j(0)=\psi_k(0)=:\psi(0)\,, \quad j,k\in\hat{n}\,, \qquad
\sum^{n}_{j=1}\psi_j'(0)=\alpha\psi(0)\,,
\end{equation}
where $\alpha\in\R$ and for brevity we have introduced the symbol
$\hat{n}:=\{1,2,\ldots,n\}$. We can add the case corresponding
formally to $\alpha=\infty$, when the system decomposes into $n$
halflines with Dirichlet endpoints, however, it is not interesting
as long as we are concerned with nontrivial vertex couplings. In
the particular case $\alpha=0$ we speak about \emph{free boundary
conditions} since for the $\delta$ function on line, $n=2$, this
corresponds to a free motion (sometimes the term Kirchhoff b.c.,
not very appropriate, is used). In terms of (\ref{parametrizace})
the $\delta$ coupling corresponds to the matrix $U= \frac{2}{n
+\i\alpha}\JJ-I$, where $\JJ$ denotes the $n\times n$ matrix whose
all entries equal one.

The $\delta'$ interaction on the line has two possible analogues
for $n>2\,$ \cite{Ex95, Ex96a}. One is a counterpart to
(\ref{delta}) called \emph{$\delta'_s$ coupling} with the role of
$\Psi(0),\, \Psi'(0)$ interchanged,
\begin{equation}\label{delta'_s}
\psi_j'(0)=\psi_k'(0)=:\psi'(0)\,, \quad j,k\in\hat{n}\,, \qquad
\sum^{n}_{j=1}\psi_j(0)=\beta\psi'(0)\,,
\end{equation}
where $\beta\in\R\cup\{+\infty\}$. It corresponds to $U=
I-\frac{2}{n-\i\beta}\JJ$, in particular, the case $\beta=\infty$
refers to full Neumann decoupling. The other one, called $\delta'$
\emph{coupling}, is
\begin{equation}\label{delta'}
\sum^{n}_{j=1}\psi_j'(0)=0\,, \qquad
\psi_j(0)-\psi_k(0)=\frac{\beta}{n}\left(\psi_j'(0)-\psi_k'(0)\right)
\quad j,k\in\hat{n}\,,
\end{equation}
with $\beta\in\R\cup\{+\infty\}$ which corresponds to $U=
-\frac{n+\i\beta}{n-\i\beta}I+ \frac{2}{n-\i\beta} \JJ$.

All the above examples have a common property, namely that the
corresponding operators are invariant with respect to permutation
of the edges, which is clear from the fact that matrices $U$ are not changed by
a simultaneous permutations of the rows and columns. The most
general family of $H_U$ with this property is characterized by two
parameters, $U=aI+bJ$ with $|a|=1$ and $|a+nb|=1$,
\emph{cf.}~\cite{ET06}, the corresponding boundary conditions
being
 \begin{eqnarray}\label{perm-sym}
 (a-1)\left(\psi_j(0)-\psi_k(0)\right)+\i(a+1)\left(\psi_j'(0)
 -\psi_k'(0)\right) &=& 0\,, \quad j,k\in\hat n\,, \nonumber \\
 [-.8em] \\ [-.8em]
 (a-1+nb)\sum^{n}_{k=1}\psi_k(0)
 +\i(a+1+nb)\sum^{n}_{k=1}\psi_k'(0)&=&0\,. \nonumber
 \end{eqnarray}

\subsection{Approximation of $\delta$ couplings}

Let us next recall briefly known results about approximations of
vertex couplings starting from the $\delta$ coupling. The idea is
the same as for $\delta$ interactions on the line.

Let $U_\delta(\alpha):= \frac{2}{n+\i\alpha}\JJ-I$ be the
corresponding matrix of the condition (\ref{parametrizace}). Given
a family of real-valued functions $W=\{W_j:\: j\in\hat n\}$, for
simplicity assumed to be compactly supported, we define scaled
potentials at graph edges by
\begin{equation} \label{scaled potentials}
W_{\epsilon,j}\,:=\, \frac{1}{\epsilon}\, W_j\left(x\over\epsilon
\right)\,,\quad j\in\hat n\,.
\end{equation}
Starting from the free boundary conditions and choosing the family
(\ref{scaled potentials}) we can approximate any nontrivial
$\delta$ coupling as the following result shows.
   \begin{theorem} \label{approximation thm}
Suppose that $W_j\in L^1(0,1)$ for $j\in\hat n$, then
\begin{equation} \label{approximation}
H_{U_\delta(0)}+W_{\epsilon}\,\longrightarrow\,
H_{U_\delta(\alpha)} \;\quad \mathrm{as} \quad \epsilon\to 0+
   \end{equation}
in the norm resolvent sense, where $\,\alpha:= \sum_{j=1}^n
\int_0^1 W_j(x)\,dx\,$.
   \end{theorem}
\noindent For \textbf{proof} see \cite{E96} where a more general
result of this type is derived, together with other extensions of
the standard Sturm-Liouville theory to star graphs.

\subsection{Approximation of singular permutation-invariant couplings}

Consider further permutation-invariant couplings with wave
functions discontinuous at the vertex. Denote the operator $H_U$
corresponding to $U=aI+bJ$ with $a,b$ satisfying the stated
conditions as $H^{a,b}$. The approximating family can be
constructed as follows: we start from the operator $H_{u,0}:=
H_{U_\delta(u)}$ and pass to $H_{u,v}$ obtained by adding a
$\delta$ interaction of strength $v$ on each edge at a distance
$d$ from the centre. We will let the $\delta$'s approach the
centre scaling properly $u,v$.

\begin{theorem} \label{singular p-i}
Fix a pair of complex numbers $a\ne -1$ and $b\ne 0$ such that
$|a|=1$ and $|a+nb|=1$, and set
\begin{equation} \label{u,v}
u(d):=\i\frac{n}{d^2} \left(\frac{a-1+nb}{a+1+nb}
 +\frac{a-1}{a+1}\right)^{-1}\,,\quad
 v(d):=-\frac{1}{d}-\i\frac{a-1}{a+1}\,.
   \end{equation}
Suppose that $a+1+nb \ne 0$ and $a(a+nb)\ne 1$, then the operators
$H_{u(d),v(d)}$ converge to $H^{a,b}$ in the norm resolvent
topology as $d\to 0+$. Moreover, the claim remains true in the two
excluded cases, provided we replace the above $u(d)$ by $-nd^{-1}$
and $\zeta d^{-\nu}$ with $\R\ni\zeta \ne 0$ and $\nu>2$,
respectively.
\end{theorem}

\noindent\textbf{Proof} can be found in \cite{ET06}, the
particular case of $\delta'_s$ coupling (\ref{delta'_s}) in which
$u(d)=-\beta d^{-2}$ and $v(d)= -d^{-1}$ was discussed in
\cite{CE04}.

\section{CS-type approximation of singular couplings}

After the preliminaries let us turn to our proper task, namely
approximations of singular couplings \emph{\`a la} Cheon and
Shigehara, i.e. by means of additional $\delta$ interactions,
properly scaled, on edges of our star graph, without the
requirement of permutation invariance.

\subsection{The class of approximable couplings}

The first question is how large is the class of operators $H_U$
which can can be treated in this way. We are going to answer it
using the technique of \cite{CS98}, i.e. looking into convergence
of the corresponding boundary conditions.

\begin{proposition} \label{admissible}
Let $\Gamma$ be a star graph with $n$ semi-infinite edges and
$\Gamma(d)$ be a graph obtained from $\Gamma$ by adding a finite
number of vertices at each edge. Consider a family
$\left\{\Gamma(d):\: d\in\R^+ \right\}$ of such graphs with the
properties that the number of the added vertices at each edge is
independent of $d$ and their distances from the centre are
$\OO(d)$ as $d\to 0_+$. Suppose that a family of functions $\Psi_d
\in W^{2,2}\left(\Gamma \setminus(\{c\}\cup V_d)\right)$, where
$c$ is the centre of $\Gamma$ and $V_d$ is the set of added
vertices, satisfies the conditions (\ref{delta}) with $d$-dependent
parameters, and that it converges to $\Psi\in W^{2,2}\left(\Gamma
\setminus\{c\}\right)$ which obeys the condition (\ref{1}) with
some $A,\,B$ satisfying the requirements (\ref{OP}). The family of
the conditions (\ref{1}) which can be obtained in this way depends
on $2n$ parameters if $n>2$, and on three parameters for $n=2$.
\end{proposition}

\begin{proof}
The $\delta$ coupling in the centre of $\Gamma$ is expressed by
the condition (\ref{delta}). Consider first $\delta$ interactions
on a halfline and look how the boundary values change when we pass
between different sites. Suppose that at a point $x$ the function
and its derivative have the right limits, and that $x+\epsilon$ is
the site of a $\delta$ interaction, then the Taylor expansion
gives
$$
\psi(x+\epsilon_-)=\psi(x_+)+\epsilon\psi'(x_+)+\OO(\epsilon^2)\,,
\quad \psi'(x+\epsilon_-)=\psi'(x_+)+\OO(\epsilon)\,,
$$
and the $\delta$ interaction is according to (\ref{delta})
described by
$$
\psi(x+\epsilon_+)=\psi(x+\epsilon_-)=:\psi(x+\epsilon)\,, \qquad
\psi'(x+\epsilon_+)-\psi'(x+\epsilon_-)=\alpha(\epsilon)\psi(x+\epsilon)\,,
$$
where $\alpha(\epsilon)$ is the coupling parameter. The may be
$\epsilon$-dependent but we suppose such a dependence that the
error terms can be neglected as $\epsilon\to 0_+$; then we have
\begin{align*}
\psi(x+\epsilon)&=\psi(x_+)+\epsilon\psi'(x_+)+\OO(\epsilon^2)\,,\\
\psi'(x+\epsilon_+)&=\psi'(x_+)+\OO(\epsilon)+\alpha(\epsilon)(\psi(x_+)
+\epsilon\psi'(x_+)+\OO(\epsilon^2))=
\\&=(1+\alpha(\epsilon)\epsilon)\psi'(x_+)+\alpha(\epsilon)\psi(x_+)
+\OO(\epsilon)+\alpha(\epsilon)\OO(\epsilon^2)\,,
\end{align*}
so that $\psi(x+\epsilon)$ and $\psi'(x+\epsilon_+)$ depend on
$\psi(x_+)$ and $\psi'(x_+)$ linearly up to error terms. In case
of a finite number of $\delta$ interactions on a halfline one can
show in a similar way recursively that the function value and the
right limit of the derivative at the site of the last $\delta$
depends, up to error terms, linearly on the function value and the
right limit of the derivative for the first $\delta$ interaction.

Let us apply this conclusion to the edges of our star graph. We
denote by $d_j$ the distance of the last $\delta$ interaction on
the $j$-th halfline family of edges in $\Gamma(d)$; by assumption
we have $d_j=\OO(d)$. Then we have
\begin{eqnarray}
\tilde{f}^{(1)}_j(d)\psi_j(d_j) &=& \tilde{g}^{(1)}_j(d)\psi(0)
+\tilde{h}^{(1)}_j(d)\psi_j'(0)+\tilde{r}^{(1)}_j(d)\,, \nonumber\\
\tilde{f}^{(2)}_j(d)\psi_j'({d_j}_+) &=&
\tilde{g}^{(2)}_j(d)\psi(0)+\tilde{h}^{(2)}_j(d)\psi_j'(0)
+\tilde{r}^{(2)}_j(d) \nonumber
\end{eqnarray}
for some $\tilde{f}^{(1)}_j, \tilde{g}^{(1)}_j, \tilde{h}^{(1)}_j,
\tilde{f}^{(2)}_j,\tilde{g}^{(2)}_j, \tilde{h}^{(2)}_j:\R^+\to\R$.
The functions $\tilde{r}^{(1)}_j$ and $\tilde{r}^{(2)}_j(d)$ are
error terms and we suppose that they can be neglected in the
limit. We are interested in the situation when the last relations
can be inverted and $\psi(0),\,\psi_j'(0)$ can be expressed by
means of $\psi_j(d_j)$ and $\psi_j'({d_j}_+)$,
\begin{eqnarray}
\psi(0) &=&
f^{(1)}_j(d)\psi_j(d_j)+g^{(1)}_j(d)\psi_j'({d_j}_+)+\RR(d)\,,
\quad j\in\hat{n}\,, \label{jhodnota} \\
\psi_j'(0) & =&
\tilde{f}^{(2)}_j(d)\psi_j(d_j)+\tilde{g}^{(2)}_j(d)\psi_j'({d_j}_+)+\RR(d)\,,
\quad\;\, j\in\hat{n}\,, \label{jderivace}
\end{eqnarray}
where we have introduced $\RR(d)$ as the symbol for a generic
remainder; we still assume that it can be neglected with respect
to the other terms as $d\to0_+$. The equations~\eqref{jhodnota}
yield for $j,k\in\hat{n}$ the conditions
\begin{equation}\label{xodectenim}
f^{(1)}_j(d)\psi_j(d_j)-f^{(1)}_k(d)\psi_k(d_k)+g^{(1)}_j(d)\psi_j'({d_j}_+)
-g^{(1)}_k(d)\psi_k'({d_k}_+)=\RR(d)\,, \quad j,k\in\hat{n}
\end{equation}
and from \eqref{jderivace} together with the second one of the
conditions \eqref{delta} we get
\begin{equation}\label{nula}
\alpha\psi(0)=\sum^{n}_{k=1}\left(f^{(2)}_k(d)\psi_k(d_k)
+g^{(2)}_k(d)\psi_k'({d_k}_+)\right)+\RR(d)\,.
\end{equation}
We substitute for $\psi(0)$ from~\eqref{jhodnota} and perform a
repeated summation of~\eqref{nula} over $j$. After an easy
rearrangement we get
\begin{equation}\label{xsectenim}
\sum^{n}_{j=1}\left(\alpha
f^{(1)}_j(d)-nf^{(2)}_j(d)\right)\psi_j(d_j)+\sum^{n}_{j=1}\left(\alpha
g^{(1)}_j(d)-ng^{(2)}_k(d)\right)\psi_j'({d_j}_+)=\RR(d)\,.
\end{equation}
Now we pass to the limit $d\to 0_+$ in the
equations~\eqref{xodectenim} and~\eqref{xsectenim}. Before that we
multiply both sides by a power of $d$ such that the right-hand
side tends to zero as $d\to 0_+$, while at least one coefficient
at the left-hand side remains nonzero, in other words, we use the
assumed existence of the limit in which the error terms can be
neglected w.r.t. the leading ones. The equation~\eqref{xodectenim}
acquires then the form
\begin{equation}\label{limodectenim}
c_j\psi_j(0)-c_k\psi_k(0)+t_j\psi_j'(0_+)-t_k\psi_k'(0_+)=0\,,
\quad j,k\in\hat{n}
\end{equation}
while~\eqref{xsectenim} gives
\begin{equation}\label{limsectenim}
\sum^{n}_{j=1}\gamma_j\psi_j(0)+\sum^{n}_{j=1}\tau_j\psi_j'(0_+)=0\,,
\end{equation}
where $c_j,\, t_j,\, \gamma_j,\, \tau_j$ are the appropriate
limiting values of the functions involved. The obtained conditions
can also be written in a matrix form,
\begin{equation}\label{OPAB}
\underbrace{\left(
\begin{array}{ccccc}
    c_1 & -c_2 & 0 & \cdots & 0 \\
    c_1 & 0 & -c_3 & \cdots & 0 \\
    \vdots & & & \ddots & \\
    c_1 & 0 & 0 & \cdots & -c_n \\
  \gamma_1 & \gamma_2 & \gamma_3 & \cdots & \gamma_n
\end{array}
\right)}_{A} \Psi(0) + \underbrace{\left(
\begin{array}{ccccc}
    t_1 & -t_2 & 0 & \cdots & 0 \\
    t_1 & 0 & -t_3 & \cdots & 0 \\
    \vdots & & & \ddots & \\
    t_1 & 0 & 0 & \cdots & -t_n \\
  \tau_1 & \tau_2 & \tau_3 & \cdots & \tau_n
\end{array}
\right)}_{B} \Psi'(0) =0\,.
\end{equation}
It is clear already now -- from the fact that the coefficients
$c_j,\, t_j,\, \gamma_j,\, \tau_j,\: j\in\hat{n}$ are real-valued
-- that the achievable number of parameters cannot exceed $4n$.

So far we have not brought the self-adjointness into the game. To
find the true number of parameters we pass from $A,\,B$ to the
unitary matrix $U$ of standard boundary conditions
(\ref{parametrizace}). This is achieved by multiplying the
relation~\eqref{OPAB} from the left by a regular matrix $M$ such
that $U-I=MA$ and $\i(U+I)=MB$. This determines $U$ since the last
relations imply
$$
U=\frac{1}{2}M(A-\i B)\,, \quad I=-\frac{1}{2}M(A+\i B)\,;
$$
notice that $A+\i B$ is regular because $A$ and $B$ are real and
the matrix $(A|B)$ has he full rank by assumption. Hence we have
$M=-2(A+\i B)^{-1}$, which further gives
$$
U=-(A+\i B)^{-1}\cdot(A-\i B)\,.
$$
We shall apply the Gauss elimination method to get the chain of
equivalences
$$
\bigl(-(A+\i B)|(A-\i B)\bigr)\sim\cdots\sim
\bigl(I|\underbrace{-(A+\i B)^{-1}\cdot(A-\i B)}_{U}\bigr)\,;
$$
the explicit form of $A\pm iB$ is obtained from (\ref{OPAB}). We
notice that the regularity of $A+iB$ implies the following facts:
(i) there is at most one $j\in\hat{n}$ such that $c_j+\i t_j=0$
(and for such a $j$ it holds that $\gamma_j+\i\tau_j\neq 0$), (ii)
there is at least one $j\in\hat{n}$ such that $\gamma_j+\i\tau_j
\neq 0$. The matrix $\bigl(-(A+\i B)|(A-\i B)\bigr)$ equals to
\begin{equation*}
\left(
\begin{array}{cccc}
    -(c_1+\i t_1) & c_2+\i t_2 & \cdots & 0 \\
    -(c_1+\i t_1) & 0 & \cdots & 0 \\
    \vdots & & \ddots & \\
    -(c_1+\i t_1) & 0 & \cdots & c_n+\i t_n \\
  -(\gamma_1+\i\tau_1) & -(\gamma_2+\i\tau_2) & \cdots & -(\gamma_n+\i\tau_n)
\end{array}
\vline
\begin{array}{cccc}
    c_1-\i t_1 & -(c_2-\i t_2) & \cdots & 0 \\
    c_1-\i t_1 & 0 & \cdots & 0 \\
    \vdots & & \ddots & \\
    c_1-\i t_1 & 0 & \cdots & -(c_n-\i t_n) \\
  \gamma_1-\i\tau_1 & \gamma_2-\i\tau_2 & \cdots & \gamma_n-\i\tau_n
\end{array}
\right)\,.
\end{equation*}
Suppose first that $c_j+\i t_j\neq 0$ for all $j\in\hat{n}$, then
by equivalent row manipulations we pass to the matrix $(D|V)$,
where
\begin{align*}
D&=\left(
\begin{array}{ccccc}
    -\left(\gamma_1+\i\tau_1+(c_1+\it_1)\sum^{n}_{\ell=1}
    \frac{\gamma_\ell+\i\tau_\ell}{c_\ell+\i t_\ell}\right)
    & 0 & 0 & \cdots & 0 \\
    0 & c_2+\i t_2 & 0 & \cdots & 0 \\
    0 & 0 & c_3+\i t_3 & \cdots & 0 \\
    \vdots & & & \ddots & \\
    0 & 0 & 0 & \cdots & c_n+\i t_n
\end{array}
\right),
\\
V&=\left(
\begin{array}{cccc}
    (c_1-\i t_1)S-2\i\frac{c_1\tau_1-\gamma_1t_1}{c_1+\i t_1}
    & -2\i\frac{c_2\tau_2-\gamma_2t_2}{c_2+\i t_2} & \cdots &
    -2\i\frac{c_n\tau_n-\gamma_n t_n}{c_n+\i t_n} \\
    \frac{2\i}{S}\frac{c_1\tau_1-\gamma_1t_1}{c_1+\i t_1} &
    -c_2+\i t_2+\frac{2\i}{S}\frac{c_2\tau_2-\gamma_2t_2}{c_2+\i t_2}
    & \cdots & \frac{2\i}{S}\frac{c_n\tau_n-\gamma_n t_n}{c_n+\i t_n} \\
    \vdots & & \ddots & \\
    \frac{2\i}{S}\frac{c_1\tau_1-\gamma_1t_1}{c_1+\i t_1}
    & \frac{2\i}{S}\frac{c_2\tau_2-\gamma_2t_2}{c_2+\i t_2}
    & \cdots & -c_n+\i t_n+\frac{2\i}{S}
    \frac{c_n\tau_n-\gamma_n t_n}{c_n+\i t_n} \\
\end{array}
\right),
\end{align*}
where we have denoted $S=\sum^{n}_{\ell=1}
\frac{\gamma_\ell+\i\tau_\ell}{c_\ell+\i t_\ell}$. Since we used
only equivalent manipulations, the diagonal matrix $D$ should have
the same rank as $A+\i B$, hence it must be regular because none
of its diagonal elements is zero. Consequently, we can divide each
row of $(D|V)$ by the corresponding diagobal element of $D$. This
yields $(I|U)$, where $U$ is the sought unitary matrix and its
diagonal and off-diagonal elements are given by
\begin{equation}\label{U}
\begin{aligned}
U_{jj}&=\frac{2\i(c_j\tau_j-t_j\gamma_j)}{(c_j+\i t_j)^2
\sum^{n}_{\ell=1}\frac{\gamma_\ell+\i\tau_\ell}{c_\ell+\i t_\ell}}
-\frac{c_j-\i t_j}{c_j+\i t_j}\,, \\
U_{jk}&=\frac{2i(c_k\tau_k-t_k\gamma_k)}{(c_j+\i t_j)(c_k+\i
t_k)\sum^{n}_{\ell=1}\frac{\gamma_\ell+\i\tau_\ell}{c_\ell+\i
t_\ell}}\quad\; \mathrm{if}\quad j\neq k\,.
\end{aligned}
\end{equation}
The right-hand sides make sense due to the first of the conditions
(\ref{OP}) and our assumptions about non-vanishing of all the
expressions $c_j+it_j$.

So far we have not employed the second one of the requirements
\eqref{OP}, namely the self-adjointness of the matrix $AB^*$. This
is equivalent to unitarity of $U$, however, it is easier to check
it in its original version. By a straightforward computation we
find that the product $AB^*=AB^T$ equals
 $$
\left(
\begin{array}{ccccccc}
    c_1t_1+c_2t_2 & c_1t_1 & c_1t_1 & \cdots & & c_1t_1  & c_1\tau_1-c_2\tau_2 \\
    c_1t_1 & c_1t_1+c_3t_3 & c_1t_1 & \cdots & & c_1t_1  & c_1\tau_1-c_3\tau_3 \\
    c_1t_1 & c_1t_1 & c_1t_1+c_4t_4 & \cdots & & c_1t_1  & c_1\tau_1-c_4\tau_4 \\
    \vdots & & & \ddots & & \vdots \\
    c_1t_1 & c_1t_1 & c_1t_1 & \cdots & & c_1t_1+c_n t_n  & c_1\tau_1-c_n\tau_n \\
    \gamma_1t_1-\gamma_2t_2 & \gamma_1t_1-\gamma_3t_3 & \gamma_1t_1-\gamma_4t_4 &
    \cdots & & \gamma_1t_1-\gamma_n t_n  & \gamma_1\tau_1+\gamma_2\tau_2+\cdots\gamma_n\tau_n
\end{array}
\right)\,,
 $$
hence $AB^*$ is self-adjoint if and only if $c_1\tau_1-c_j\tau_j
=\gamma_1t_1-\gamma_j t_j$ holds for all $j=2,\ldots,n$, and
therefore
\begin{equation}\label{podmunit}
c_1\tau_1-\gamma_1t_1=c_2\tau_2-\gamma_2t_2=c_3\tau_3-\gamma_3t_3=\cdots
=c_n\tau_n-\gamma_n t_n\,.
\end{equation}
We denote the common value $c_j\tau_j-\gamma_j t_j$ as $\kappa$
and recall that we have denoted $S=\sum^{n}_{\ell=1}
\frac{\gamma_\ell+\i\tau_\ell}{c_\ell+\i t_\ell}$, then the matrix
$U$ given by~\eqref{U} can be simplified,
$$
U=\left(
\begin{array}{cccc}
    \frac{2\i\kappa}{(c_1+\i t_1)^2S}-\frac{c_1-\i t_1}{c_1+\i t_1} &
    \frac{2\i\kappa}{(c_1+\i t_1)(c_2+\i t_2)S} & \cdots &
    \frac{2\i\kappa}{(c_1+\i t_1)(c_n+\i t_n)S} \\
    \frac{2\i\kappa}{(c_2+\i t_2)(c_1+\i t_1)S} &
    \frac{2\i\kappa}{(c_2+\i t_2)^2S}-\frac{c_2-\i t_2}{c_2+\i t_2} &
    \cdots & \frac{2\i\kappa}{(c_2+\i t_2)(c_n+\i t_n)S} \\
    \vdots & & \ddots & \\
    \frac{2\i\kappa}{(c_n+\i t_n)(c_1+\i t_1)S} &
    \frac{2\i\kappa}{(c_n+\i t_n)(c_2+\i t_2)S} &
    \cdots & \frac{2\i\kappa}{(c_n+\i t_n)(c_n+\i t_n)S}
    -\frac{c_n-\i t_n}{c_n+\i t_n}
\end{array}
\right)\,.
$$
Let us show that the matrix ~\eqref{U} can be parametrized by $2n$
real numbers. We rewrite the quantity $S$ introduced above in the
following way,
 $$
 S=\sum^{n}_{\ell=1}\frac{(\gamma_\ell+\i\tau_\ell)
 (c_\ell-\i t_\ell)} {c_\ell^2+t_\ell^2}
 =\sum^{n}_{\ell=1}\frac{c_\ell\gamma_\ell+t_\ell\tau_\ell}
 {c_\ell^2+t_\ell^2}
 +\i\kappa\sum^{n}_{\ell=1}\frac{1}{c_\ell^2+t_\ell^2}\,,
 $$
and make first several observations: (i) regarding~\eqref{OPAB} as
a system of linear equations its solvability is not affected if
the last one is multiplied by a nonzero number. At the same time,
the value of $\kappa$ is directly proportional to $\gamma_j$,
$\tau_j$, and consequently, one can suppose without loss of
generality that $\kappa=1\,$ (the case $\kappa=0$ gives rise to
the same situation as $c_1+\i t_1=0$ which we shall discuss
below), (ii) if $\kappa=1$ the imaginary part of $S$ is determined
only by the values of $c_j,\,t_j,\, j\in\hat n\,$, (iii) and
finally, one can also suppose without loss of generality that
$|c_1+\i t_1|=1$, since in the opposite case we can divide all but
the last of the equations in the system~\eqref{OPAB} by $|c_1+\i
t_1|$ which is nonzero by assumption.

With the above convention we can denote $c_1+\i t_1=:
\e^{\i\theta}$ and $\Re S=:\rho$ so that
 $$
 S=\rho+\i\left(1+\sum^{n}_{\ell=2}\frac{1}{c_\ell^2+t_\ell^2}\right)
 $$
and $U$ can be written explicitly as
\begin{equation}\label{hezkaU}
U=\left(
\begin{array}{cccc}
\frac{2\i}{S}\ \e^{-2\i\theta}-\e^{-2\i\theta} & \frac{2\i}
{(c_2+\i t_2)S}\ \e^{-\i\theta} &
\cdots & \frac{2\i}{(c_n+\i t_n)S}\ \e^{-\i\theta} \\
    \frac{2\i}{(c_2+\i t_2)S}\ \e^{-\i\theta} &
    \frac{2\i}{(c_2+\i t_2)^2S}-\frac{c_2-\i t_2}{c_2+\i t_2} &
    \cdots & \frac{2\i}{(c_2+\i t_2)(c_n+\i t_n)S} \\
    \vdots & & \ddots & \\
    \frac{2\i}{(c_n+\i t_n)S}\ \e^{-\i\theta} &
    \frac{2\i}{(c_n+\i t_n)(c_2+\i t_2)S} & \cdots &
    \frac{2\i}{(c_n+\i t_n)(c_n+\i t_n)S}-\frac{c_n-\i t_n}{c_n+\i t_n}
\end{array}
\right)\,.
\end{equation}
being dependent on $2n$ real parameters $\theta,c_2,c_3,\ldots,
c_n,t_2,t_3,\ldots t_n,\rho$.

The above argument applies to any $n>2$. In the case $n=2$ the
situation is somewhat different, because we have $n^2=2n=4$
but~\eqref{hezkaU} does not give the whole family of unitary
$2\times 2$ matrices; notice that the off-diagonal elements
coincide. It is easy to show that the admissible $U$ can be for
$n=2$ characterized by three real parameters. Indeed, writing $U=
{a\;b \choose b\;c}$ the unitarity requirement reads
 $$
|a|^2+|b|^2=1,\, \quad |b|^2+|c|^2=1,\, \quad
a\bar{b}+b\bar{c}=0\,.
 $$
Knowing the modulus and phase of $a$, the modulus of $b$ is
determined so one has to choose its phase. Furthermore, since we
assume $b\neq0$ the element $c$ is uniquely determined. Hence the
matrix $U$ of~\eqref{hezkaU} is described by three parameters
which can be chosen, e.g., as the real parts of $U_{jj}$ and the
phase of $U_{12}$.

Returning to the general case one can also write the conditions
(\ref{1}) explicitly in terms of the parameters. A straightforward
way is to put $\tilde{A}=U-I,\,\tilde{B}=\i(U+I)$ with $U$ given
by~\eqref{hezkaU}. To get a simpler expression one can pass from
the system $\tilde{A}\Psi(0)+\tilde{B}\Psi'(0)=0$ to an equivalent
one multiplying it from the left by the matrix
$$
\frac{1}{2}\left(
\begin{array}{ccccc}
    -\e^{\i\theta} & c_2+\i t_2 & 0 & \cdots & 0 \\
    -\e^{\i\theta} & 0 & c_3+\i t_3 & \cdots & 0 \\
    \vdots & & & \ddots & \\
    -\e^{\i\theta} & 0 & 0 & \cdots & c_n+\i t_n \\
    \e^{\i\theta} & c_2+\i t_2 & c_3+\i t_3 & \cdots & c_n+\i t_n
\end{array}
\right)\,;
$$
this yields an explicit parametrization of the conditions
(\ref{1}) with
\begin{equation}\label{hezkeOP}
\begin{aligned}
A&= \left(
\begin{array}{ccccc}
    \cos\theta & -c_2 & 0 & \cdots & 0 \\
    \cos\theta & 0 & -c_3 & \cdots & 0 \\
    \vdots & & & \ddots & \\
    \cos\theta & 0 & 0 & \cdots & -c_n \\
  S\cos\theta-\frac{\i}{c_1+\i t_1} & Sc_2-\frac{\i}{c_2+\i t_2} &
  Sc_3-\frac{\i}{c_3+\i t_3} & \cdots & Sc_n-\frac{\i}{c_n+\i t_n}
\end{array}
\right)\,,\\[.5em]
B&= \left(
\begin{array}{ccccc}
    \sin\theta & -t_2 & 0 & \cdots & 0 \\
    \sin\theta & 0 & -t_3 & \cdots & 0 \\
    \vdots & & & \ddots & \\
    \sin\theta & 0 & 0 & \cdots & -t_n \\
  S\sin\theta+\frac{1}{c_1+\i t_1} & St_2+\frac{1}{c_2+\i t_2} &
  St_3+\frac{1}{c_3+\i t_3} & \cdots & St_n+\frac{1}{c_n+\i t_n}
\end{array}
\right)
\end{aligned}
\end{equation}
and concludes the argument in the generic case when $c_j+\i
t_j\neq 0$ for all $j\in\hat{n}$.

It remains to deal with the case when the last mentioned
requirement is violated; without loss of generality we may suppose
that $c_1+\i t_1=0$. The matrix $\bigl(-(A+\i B)|(A-\i B)\bigr)$
has then the form
\begin{equation*}
\left(
\begin{array}{cccc}
    0 & c_2+\i t_2 & \cdots & 0 \\
    \vdots & & \ddots & \\
    0 & 0 & \cdots & c_n+\i t_n \\
  -(\gamma_1+\i\tau_1) & -(\gamma_2+\i\tau_2) & \cdots & -(\gamma_n+\i\tau_n)
\end{array}
\vline
\begin{array}{cccc}
    0 & -(c_2-\i t_2) & \cdots & 0 \\
    \vdots & & \ddots & \\
    0 & 0 & \cdots & -(c_n-\i t_n) \\
  \gamma_1-\i\tau_1 & \gamma_2-\i\tau_2 & \cdots & \gamma_n-\i\tau_n
\end{array}
\right)
\end{equation*}
Using the Gauss elimination scheme we arrive at $(D|V)$ with a
diagonal $D$ and upper-triangular $V$, and from here in the same
way as above to $(I|U)$ with
\begin{equation*} 
U=\left(
\begin{array}{ccccc}
  -\frac{\gamma_1-\i\tau_1}{\gamma_1+\i\tau_1} &
  \frac{2\i}{\gamma_1+\i\tau_1}\frac{c_2\tau_2-\gamma_2t_2}{c_2+\i t_2} &
  \frac{2\i}{\gamma_1+\i\tau_1}\frac{c_3\tau_3-\gamma_3t_3}{c_3+\i t_3} &
  \cdots & \frac{2\i}{\gamma_1+\i\tau_1}\frac{c_n\tau_n-\gamma_n t_n}{c_n+\i t_n} \\
    0 & -\frac{c_2-\i t_2}{c_2+\i t_2} & 0 & \cdots & 0 \\
    0 & 0 & -\frac{c_3-\i t_3}{c_3+\i t_3} & \cdots & 0 \\
    \vdots & & & \ddots & \\
    0 & 0 & 0 & \cdots & -\frac{c_n-\i t_n}{c_n+\i t_n}
\end{array}
\right)\,.
\end{equation*}
Furthermore, it follows from the condition~\eqref{podmunit} with
$c_1=t_1=0$ that
$$
c_2\tau_2-\gamma_2t_2=c_3\tau_3-\gamma_3t_3=\cdots
=c_n\tau_n-\gamma_n t_n=0\,.
$$
hence all the off-diagonal elements in the above matrix $U$ vanish
which means that it is characterized by $n$ real parameters,
$$
U=\mathrm{diag}\left\{ \e^{\i\theta_1}, \ldots, \e^{\i\theta_n}
\right\}\,.
$$
It is easy to rewrite the boundary conditions in the form
(\ref{1}) and check that they correspond to the fully separated
case,
\begin{equation}\label{npar}
\sin\frac{\theta_j}{2}\cdot\psi_j(0)+\cos\frac{\theta_j}{2}\cdot\psi'_j(0)=0\,,
\quad j\in\hat{n}\,,
\end{equation}
which is, of course, trivial for the viewpoint of quantum
mechanics on $\Gamma$. \end{proof}

\subsection{A concrete $2n$-parameter approximation}

Knowing the maximum number of parameters in the boundary
conditions (\ref{1}) which can be achieved in this way, we are
naturally lead to the idea of placing two $\delta$ interactions at
each of the $n$ halflines. In this section we are going to
concretize this proposal. We will concentrate at the
matrix~\eqref{hezkaU} in the generic case leaving out the trivial
situation (\ref{npar}) mentioned at the end of the previous proof.
We will also leave out the case $n=2$ which was discussed in the
paper~\cite{SMMC99}.

\begin{figure}[h]
\begin{center}
\includegraphics[angle=0,width=0.5\textwidth]{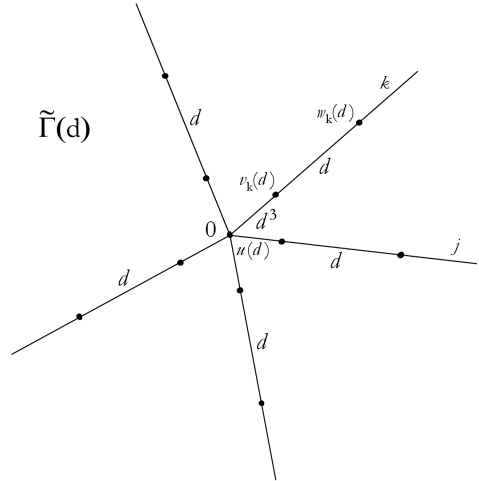}
\caption{Scheme of a $2n$-parameter approximation} \label{2int}
\end{center}
\end{figure}

Let us specify the approximation arrangement. The $\delta$'s are
placed as sketched on~Fig.~\ref{2int}, all dependent on a
parameter $d$ in terms of which the limit is performed:
 \begin{itemize}
 \item there is a $\delta$ coupling with parameter $u(d)$ in the
 star centre
 \item on each halfline there is a $\delta$ interaction with parameter
 $v_j(d)$, where $j$ is the halfline index, at a distance $D(d)$
 from the centre (it will turn out in the following that we may choose $D(d)=d^3$)
 \item furthermore, each halfline supports another $\delta$ interaction
 with parameter $w_j(d)$ at the distance $D(d)+d$ from the centre
\end{itemize}
For the sake of brevity we will not indicate the $d$-dependence of
the parameters $u,\, v_j,\, w_j$ and the distance $D$ unless
necessary. The boundary conditions which the functions
$\psi_1,\ldots,\psi_n$ on $\Gamma$ have to satisfy are
\begin{eqnarray}
 && \psi_1(0)=\psi_2(0)=\cdots=\psi_n(0)=:\psi(0)\,, \qquad
\sum^{n}_{j=1}\psi_j'(0_+)=u\psi(0) \label{I.serie2n} \\
&& \psi_j(D_+)=\psi_j(D_-)=:\psi_j(D)\,,\quad
\psi_j'(D_+)-\psi_j'(D_-)=v_j\psi_j(D) \label{IIv.serie2n} \\
&& \psi_j(D+d_{\pm})=:\psi_j(D+d)\,,\quad
\psi_j'(D+d_+)-\psi_j'(D+d_-)=w_j\psi_j(D+d)\,. \phantom{A}
\label{IIw.serie2n}
\end{eqnarray}
Further relations which will in the following serve to determine
the parameter dependence on $d$ are obtained from Taylor expansion
of the respective wave functions,
\begin{eqnarray}
 && \psi_j(D)=\psi_j(0)+D\psi_j'(0_+)+\OO(D^2)\,, \qquad
\psi_j'(D_-)=\psi_j'(0_+)+\OO(D)\,, \label{IIIv.serie2n} \\
&& \psi_j(D+d)=\psi_j(D)+d\psi_j'(D_+)+\OO(d^2)\,, \;
\psi_j'(D+d_-)=\psi_j'(D_+)+\OO(d) \phantom{AA}
\label{IIIw.serie2n}
\end{eqnarray}
for $j\in\hat{n}$. We need to find relations between the values
$\psi_1(D+d),\ldots, \psi_n(D+d)$ and $\psi_1'(D+d_+),
\ldots,\psi_n'(D+d_+)$. To this aim we express them first in terms
of $\psi(0)$ and $\psi_j'(0_+)$. Using the
relations~\eqref{IIv.serie2n} and~\eqref{IIIv.serie2n} we get
\begin{align*}
\psi_j'(D_+)&=\psi_j'(0_+)+\OO(D)+v_j(\psi_j(0)+D\psi_j'(0_+)+\OO(D^2))=\\
&=v_j\psi(0)+(1+v_jD)\psi_j'(0_+)+\OO(D)+v_j\OO(D^2)\,.
\end{align*}
Substituting into the first one of the
relations~\eqref{IIIw.serie2n} and using ~\eqref{IIv.serie2n}
again we find
\begin{multline}\label{psi(D+d)}
\psi_j(D+d) =(1+dv_j)\psi(0)
+\left(D+d(1+v_jD)\right)\psi_j'(0_+) \\
+\OO(D^2)+d\OO(D)+dv_j\OO(D^2)+\OO(d^2)\,.
\end{multline}
The already obtained expression for $\psi_j'(D_+)$ together with
the second one of the relations ~\eqref{IIIw.serie2n} give
$$
\psi_j'(D+d_-)=v_j\psi(0)+(1+v_jD)\psi_j'(0_+)
+\OO(D)+v_j\OO(D^2)+\OO(d)\,.
$$
Substituting from here and~\eqref{psi(D+d)} into the second one of
the relations~\eqref{IIw.serie2n} we get after a simple
rearrangement
\begin{multline}\label{psi'(D+d+)}
\psi_j'(D+d_+)=\left(v_j+w_j(1+dv_j)\right)\psi(0)
+\left(1+v_jD+w_j\left(D+d(1+v_jD)\right)\right)\psi_j'(0_+) \phantom{A}\\
+\OO(D)+v_j\OO(D^2)+\OO(d)+w_j\left(\OO(D^2)+d\OO(D)+dv_j\OO(D^2)
+\OO(d^2)\right).
\end{multline}
Next we eliminate $\psi_j'(0_+)$ (for simplicity we write $\psi_j'(0)$) from the obtained relations~\eqref{psi(D+d)} and~\eqref{psi'(D+d+)},
multiplying them by $1+v_jD+w_j\left(D+d(1+v_jD)\right)$ and
$D+d(1+v_jD)$, respectively, and subtracting. In the resulting
expression the coefficient at $\psi(0)$ equals one,
\begin{multline}\label{elim psi'(0)}
\left(1+v_jD+w_j\left(D+d(1+v_jD)\right)\right)\psi_j(D+d) \\
=\psi(0)+\left(D+d(1+v_jD)\right)\psi_j'(D+d_+)+\RR_j\,,
\end{multline}
with the remainder term
\begin{multline*}
\RR_j:=\left(1+v_jD+w_j\left(D+d(1+v_jD)\right)\right)
(\OO(D^2)+d\OO(D)+dv_j\OO(D^2)+\OO(d^2))\\
-\left(D+d(1+v_jD)\right)\big((\OO(D)+v_j\OO(D^2)+\OO(d)\\
+w_j\left(\OO(D^2)+d\OO(D)+dv_j\OO(D^2)+\OO(d^2)\right)\big)\,.
\end{multline*}
So far the edge index has been kept fixed. Subtracting mutually
the relations \eqref{elim psi'(0)} for different values of
$j,k\in\N$, we can eliminate $\psi(0)$,
\begin{multline}\label{odectenim2n}
\left(1+v_jD+w_j\left(D+d(1+v_jD)\right)\right)\psi_j(D+d)\\
-\left(1+v_kD+w_k\left(D+d(1+v_kD)\right)\right)\psi_k(D+d)\\
=\left(D+d(1+v_jD)\right)\psi_j'(D+d_+)
-\left(D+d(1+v_kD)\right)\psi_k'(D+d_+)+\RR_j-\RR_k
\end{multline}
Returning to the relations~\eqref{psi(D+d)} and~\eqref{psi'(D+d+)}
we can eliminate from them $\psi(0)$ in a similar way as above
arriving at the relation
\begin{equation}
(1+dv_j)\psi_j'(D+d_+)-\left(v_j+w_j(1+dv_j)\right)\psi_j(D+d)
=\psi_j'(0)-\tilde{\RR}_j
\end{equation}
with the remainder term
\begin{multline*}
\tilde{\RR}_j:=\left(v_j+w_j(1+dv_j)\right)(\OO(D^2)
+d\OO(D)+dv_j\OO(D^2)+\OO(d^2))-\\
-(1+dv_j)\left((\OO(D)+v_j\OO(D^2)+\OO(d)+w_j\left(\OO(D^2)
+d\OO(D)+dv_j\OO(D^2)+\OO(d^2)\right)\right)\,.
\end{multline*}
Summing the above relations over $j\in\N$ and using
~\eqref{I.serie2n} we get
\begin{equation}\label{elim psi(0)}
\sum^{n}_{j=1}(1+dv_j)\psi_j'(D+d_+)
-\sum^{n}_{j=1}\left(v_j+w_j(1+dv_j)\right)\psi_j(D+d)
=u\psi(0)-\sum^{n}_{j=1}\tilde{\RR}_j\,.
\end{equation}
The right-hand side can rewritten using the continuity
condition~\eqref{I.serie2n} in combination with the
relation~\eqref{elim psi'(0)},
\begin{multline*}
u\psi(0)=\frac{u}{n}\sum^{n}_{j=1}\psi_j(0)=
\frac{u}{n}\sum^{n}_{j=1}\big(\left(1+v_jD+w_j
\left(D+d(1+v_jD)\right)\right)\psi_j(D+d)\\
-\left(D+d(1+v_jD)\right)\psi_j'(D+d_+)+\RR_j\big)\,.
\end{multline*}
This allows us to cast \eqref{elim psi(0)} into a form which
contains neither $\psi(0)$ nor $\psi_j'(0)$,
\begin{multline}\label{sectenim2n}
\sum^{n}_{j=1}\left(v_j+w_j(1+dv_j)+\frac{u}{n}\left(1+v_jD+w_j\left(D+d(1+v_jD)\right)\right)\right)\psi_j(D+d)=\\
=\sum^{n}_{j=1}\left(1+dv_j+\frac{u}{n}\left(D+d(1+v_jD)\right)\right)\psi_j'(D+d_+)+\sum^{n}_{j=1}\left(\tilde{\RR}_j-\frac{u}{n}\RR_j\right)\,.
\end{multline}
The equations~\eqref{odectenim2n} and~\eqref{sectenim2n} are the
sought relations between the function values and derivatives at
the sites of the ``outer'' $\delta$'s with $\psi(0)$ and
$\psi_j'(0)$ eliminated.

In the next step we are going to choose the dependences $D=D(d),\,
u=u(d),\, v_j=v_j(d)$ and $w_j=w_j(d)$ for $j\in\hat n$ in such a
way that the limit $d\to 0_+$ will yield the ($2n$-parameter
family of) boundary conditions~\eqref{1} satisfying the
requirement~\eqref{OP}. It appears that a suitable choice is the
following one,
\begin{equation}\label{u,v,w}
\begin{split}
D(d)&:=d^3 \\
1+v_jD&=\alpha_jd\,,\quad \text{i.e.} \quad
v_j(d):=-\frac{1}{d^3}+\frac{\alpha_j}{d^2} \\
1+w_jd&=\beta_jd\,,\quad \text{i.e.} \quad
w_j(d):=-\frac{1}{d}+\beta_j \\
u(d)&:=\frac{\omega}{d^4}
\end{split}
\end{equation}
Indeed, in such a case the coefficients in~\eqref{odectenim2n}
acquire the form
\begin{equation}\label{koefodect}
\begin{split}
(1+v_jD)(1+w_jd)+w_jD &=(\alpha_j\beta_j-1)d^2+\beta_jd^3\,, \\
D+d(1+v_jD)&=\alpha_jd^2+d^3
\end{split}
\end{equation}
and a straightforward computation shows that the remainders are
$\RR_j=d^2\OO(d)$, hence dividing~\eqref{odectenim2n} by $d^2$ we
arrive at
\begin{equation*}
\begin{split}
(\alpha_j\beta_j-1+\beta_jd)\psi_j(d^3+d)
-(\alpha_k\beta_k-1+\beta_kd)\psi_k(d^3+d)\\
=(\alpha_j+d)\psi_j'(d^3+d_+)
-(\alpha_k+d)\psi_k'(d^3+d_+)+\OO(d)\,.
\end{split}
\end{equation*}
Taking the limit $d\to 0_+$ we have to realize that the condition
$\psi_j\in W^{2,2}(\R^+),\, j\in\hat n$, requires that $\psi_j(d)
=o(d^{-1/2})$ holds at the halfline endpoint, hence we have
\begin{equation}\label{odectenim2nlim}
(\alpha_j\beta_j-1)\psi_j(0)-(\alpha_k\beta_k-1)\psi_k(0)
=\alpha_j\psi_j'(0)-\alpha_k\psi_k'(0)\,,\quad
j,k\in\hat{n}\,.
\end{equation}
In a similar way we proceed with the equation~\eqref{sectenim2n}.
We employ~\eqref{koefodect}, then a straightforward computation
gives for the coefficients at $\psi_j(D+d)$ and $\psi_j'(D+d_+)$
the following expressions
\begin{eqnarray*}
&& v_j+w_j(1+dv_j)+\frac{u}{n}\left(1+v_jD+w_j
\left(D+d(1+v_jD)\right)\right)\\
&& \qquad =\left(-\beta_j
+\frac{\omega}{n}(\alpha_j\beta_j-1)\right)\frac{1}{d^2}
+\left(\alpha_j\beta_j-1
+\frac{\omega}{n}\beta_j\right)\frac{1}{d}+\beta_j\,, \\
&& 1+dv_j+\frac{u}{n}\left(D+d(1+v_jD)\right)
=\left(-1+\frac{\omega}{n}\alpha_j\right)\frac{1}{d^2}
+\left(\alpha_j+\frac{\omega}{n}\right)\frac{1}{d}+1\,,
\end{eqnarray*}
and the remainder terms $\tilde{\RR}_j$ and $\frac{u}{n}\RR_j$ are
both $d^{-2}\OO(d)$. We substitute from here
to~\eqref{sectenim2n}, multiply the result by $d^2$ and pass to
the limit $d\to 0_+$; this yields
\begin{equation}\label{sectenim2nlim}
\sum^{n}_{j=1}\left(-\beta_j
+\frac{\omega}{n}(\alpha_j\beta_j-1)\right)\psi_j(0)
=\sum^{n}_{j=1}\left(-1+\frac{\omega}{n}
\alpha_j\right)\psi_j'(0)\,,\quad j\in\hat{n}\,.
\end{equation}

The relations~\eqref{odectenim2nlim} and~\eqref{sectenim2nlim} are
the sought boundary conditions. It remains to express them as
(\ref{1}) and to find relations between the parameters contained
in them to those of~\eqref{hezkaU}. The matrix form
of~\eqref{odectenim2nlim} and~\eqref{sectenim2nlim} looks as
follows,
\begin{equation}\label{navrhOP}
\left(
\begin{array}{cccc}
    \alpha_1\beta_1-1 & -(\alpha_2\beta_2-1) & \cdots & 0 \\
    \vdots & & \ddots & \\
    \alpha_1\beta_1-1 & 0 & \cdots & -(\alpha_n\beta_n-1) \\
  \tilde{\gamma}_1 & \tilde{\gamma}_2 & \cdots & \tilde{\gamma}_n
\end{array}
\right) \Psi(0) + \left(
\begin{array}{cccc}
    -\alpha_1 & \alpha_2 & \cdots & 0 \\
    \vdots & & \ddots & \\
    -\alpha_1 & 0 & \cdots & \alpha_n \\
  \tilde{\tau}_1 & \tilde{\tau}_2 & \cdots & \tilde{\tau}_n
\end{array}
\right) \Psi'(0) =0\,,
\end{equation}
where $\tilde{\gamma}_j :=\frac{\omega}{n} (\alpha_j\beta_j-1)
-\beta_j$ and $\tilde{\tau}_j:=1-\frac{\omega}{n}\alpha_j$. We
know that the corresponding matrix of (\ref{parametrizace}) is
given by $U=-(A+\i B)^{-1}\cdot(A-\i B)$, its matrix element being
\begin{align*}
U_{jj}=&\frac{2\i}{(\alpha_j\beta_j-1-\i\alpha_j)^2
\left(\sum^{n}_{l=1}\frac{\beta_l(\alpha_l\beta_l-1)
+\alpha_l}{(\alpha_l\beta_l-1)^2+\alpha_l^2}
-\omega+\i\sum^{n}_{l=1}\frac{1}{(\alpha_l\beta_l-1)^2+\alpha_l^2}\right)}
\\ & \quad
-\frac{\alpha_j\beta_j-1+\i\alpha_j}{\alpha_j\beta_j-1-\i\alpha_j}
\end{align*}
and
 $$
U_{jk}=\frac{2\i}{(\alpha_j\beta_j-1-\i\alpha_j)
(\alpha_k\beta_k-1-\i\alpha_k)
\left(\sum^{n}_{l=1}\frac{\beta_l(\alpha_l\beta_l-1)
+\alpha_l}{(\alpha_l\beta_l-1)^2+\alpha_l^2}
-\omega+\i\sum^{n}_{l=1}\frac{1}{(\alpha_l\beta_l-1)^2+\alpha_l^2}\right)}
 $$
for $j\neq k$. If the latter should correspond to ~\eqref{hezkaU},
it is sufficient to require
\begin{equation}\label{jednotka}
|\alpha_1\beta_1-1-\i\alpha_1|=1
\end{equation}
and to set
\begin{eqnarray}
&&\sum^{n}_{l=1}\frac{\beta_l(\alpha_l\beta_l-1)
+\alpha_l}{(\alpha_l\beta_l-1)^2+\alpha_l^2}-\omega=\rho\,,
\label{omegaro} \\ && \alpha_j \beta_j-1=c_j\,, \qquad
-\alpha_j=t_j\,. \label{second}
\end{eqnarray}
For $\alpha_1=0$ the condition~\eqref{jednotka} is satified
trivially, while for a nonzero value it is equivalent to $\alpha_1
\left(\alpha_1(\beta_1^2+1)-2\beta_1\right)=0$, in other words we
have to put
$$
\alpha_1=\frac{2\beta_1}{\beta_1^2+1}\,.
$$
In this way we have eliminated the parameter $\alpha_1$, and just
$2n$ of them is left. The correspondence between the $2n$-tuples
$\beta_1,\beta_2,\beta_3,\ldots \beta_n,\alpha_2,\alpha_3,\ldots,
\alpha_n,\omega$ and $\theta,c_2,c_3,\ldots c_n,t_2,t_3,\ldots,
t_n,\rho$ looks as follows:
\begin{itemize}
\item $\beta_1\longleftrightarrow\theta$: they are related by
$\frac{\beta_1-\i}{\beta_1+\i}=\e^{\i\theta}$
\item $\alpha_j,\beta_j\longleftrightarrow c_j,t_j$,
$j\in\{2,\ldots,n\}$: see (\ref{second}),
\item $\omega\longleftrightarrow\rho$: see~\eqref{omegaro}.
\end{itemize}
In what follows we will work with $\beta_1,\beta_2,\beta_3,\ldots
\beta_n,\alpha_2,\alpha_3,\ldots \alpha_n,\omega$, for simplicity
we will use also $\alpha_1$ remembering that it is determined by
$\beta_1$ and the relation~\eqref{jednotka}.

\section{Norm-resolvent convergence}

The approximation worked out in the previous section was in the
spirit of \cite{CS98, SMMC99} being expressed in terms of boundary
conditions. One asks naturally what can be said about the relation
between the corresponding operators. We denote the Hamiltonian
with the coupling (\ref{navrhOP}) in centre of the star as
$H^{\omega,\vec{\alpha},\vec{\beta}}$, and
$H^{u,\vec{v},\vec{w}}(d)$ will be the approximating family
constructed above, with a pair of $\delta$ interactions added at
each halfline. Our aim here is to demonstrate the following claim.

\begin{theorem}
Let $u,\,v_j,\,w_j,\; j\in\hat{n}$, depend on $d$ according
to~\eqref{u,v,w}, i.e.
$$
u(d)=\frac{\omega}{d^4}\,, \quad
v_j(d)=-\frac{1}{d^3}+\frac{\alpha_j}{d^2}\,, \quad
w_j(d)=-\frac{1}{d}+\beta_j\,.
$$
Then $H^{u,\vec{v},\vec{w}}(d)$ converges to
$H^{\omega,\vec{\alpha},\vec{\beta}}$ in the norm-resolvent sense
as $d\to 0_+$.
\end{theorem}

\begin{proof} We have to compare the resolvents
$R_{H^{u,\vec{v},\vec{w}}(d)}(k^2)$ and $R_{H^{\omega,
\vec{\alpha},\vec{\beta}}}(k^2)$ of the two operators for $k^2$ in
the resolvent set. It is clearly sufficient to check the
convergence in the Hilbert-Schmidt norm,
$$
\left\|R_{H^{u,\vec{v},\vec{w}}(d)}(k^2)
-R_{H^{\omega,\vec{\alpha},\vec{\beta}}}(k^2)\right\|_2\to0_+
\quad \text{ as }\: d\longrightarrow0_+\,,
$$
in other words, to show that the difference of the corresponding
resolvent kernels denoted as $\GG^{u,\vec{v}, \vec{w}}_k$ and
$\GG^{\omega, \vec{\alpha}, \vec{\beta}}_k$, respectively, tends
to zero in $L^2((\R^+)^{2n})$. Recall that these jernels, or Green
functions, are in our case $n\times n$ matrix functions.

Let us construct first $\GG^{\omega,\vec{\alpha},\vec{\beta}}_k$
for the star-graph Hamiltonian referring to the condition
(\ref{1}) in the centre. We begin with $n$ independent halflines
with Dirichlet condition at its endpoints; Green's function for
each of them is well-known to be
$$
\GG_{\i\kappa}(x,y)=\frac{\sinh\kappa x_<\: \e^{-\kappa
x_>}}{\kappa}\,,
$$
where $x_<:=\min\{x,y\},\:x_>:=\max\{x,y\}$, and we put
$\i\kappa=k$ assuming $\Re\kappa>0$. The sought Green's function
is then given by Krein's formula \cite[App.~A]{AGHH},
\begin{equation}\label{Krein1}
R_{H^{A,B}}(k^2)=R_{H}(k^2) +\sum^{n}_{j,l=1}\lambda_{jl}(k^2)
\left(\phi_l\left(\overline{k^2}\right),\cdot\right)_{
L^2((\mathbb{R}^+)^n)}\phi_j(k^2)\,,
\end{equation}
where $R_{H}(k^2)$ acts on each halfline as an integral operator
with the kernel $\GG_\kappa$ and for $\phi_j(k^2)$ one can choose
any elements of the deficiency subspaces of the largest common
restriction; we will work with $\left(\phi_j(k^2)(x)\right)_m
=\delta_{jm}\e^{-\kappa x}$.

To find the coefficients $\lambda_{jl}(k^2)$ we apply
~\eqref{Krein1} to an arbitrary $\Psi\in \bigoplus_{j=1}^n
L^2(\mathbb{R}^+)$ and denote the components of the resulting
vector as $h_j$; it yields
$$
h_j(x_j)=\int^{+\infty}_{0}\GG_{\i\kappa}(x,y_j)\psi_j(y_j)\D
y_j+\sum^{n}_{l=1}\lambda_{jl}(k^2)\int^{+\infty}_{0}\\e^{-\kappa
y_l} \psi_l(y_l)\D y_l\cdot\e^{-\kappa x_j}\,.
$$
These functions have to satisfy the boundary conditions in the
centre,
\begin{equation}\label{KrPodm}
\sum^{n}_{m=1}A_{jm}h_m(0)+\sum^{n}_{m=1}B_{jm}h_m'(0)=0 \quad
\text{for all }\; j\in\hat{n}\,.
\end{equation}
Using the explicit form of $\GG_{\i\kappa}(x,y)$ and
$\left.\frac{\partial\GG_\kappa(x_m,y_m)}{\partial
x_m}\right|_{x_m=0}=\e^{-\kappa y_m}$ we find
\begin{equation}\label{h0}
h_m(0)=\sum^{n}_{l=1}\lambda_{ml}(k^2)\int^{+\infty}_{0}\e^{-\kappa
y_l}\psi_l(y_l)\D y_l
\end{equation}
and
\begin{equation}\label{h'0}
h_m'(0)=\int^{+\infty}_{0}\e^{-\kappa y_m}\psi_m(y_m)\D
y_m-\kappa\sum^{n}_{l=1}\lambda_{ml}(k^2)\int^{+\infty}_{0}\e^{-\kappa
y_l}\psi_l(y_l)\D y_l\,.
\end{equation}
Substituting from these relations into~\eqref{KrPodm} we get a
system of equations,
$$
\sum^{n}_{l=1}\int^{+\infty}_{0}
\left(\sum^{n}_{m=1}A_{jm}\lambda_{ml}(k^2)+B_{jl}
-\kappa\sum^{n}_{m=1}B_{jm}\lambda_{ml}(k^2)\right) \e^{-\kappa
y_l}\psi_l(y_l)\D y_l=0\,,
$$
with $j\in\hat{n}$. We require that the left-hand side vanishes
for any $\psi_1,\psi_2,\ldots,\psi_n$; this yields the condition
$A\Lambda+B-\kappa B\Lambda=0$. From here it is easy to find the
coefficients $\lambda_{jl}(k^2)$: we have $(A-\kappa B)
\Lambda=-B$, and therefore
$$
\lambda_{jl}(k^2)=-\left[(A-\kappa B)^{-1}B\right]_{jl}\,.
$$
Notice that the matrix $A-\kappa B$ is regular in view of the
first conditions in~\eqref{OP}; since $A,\,B$ are real and
$\Im\kappa\neq0$, the requirement $\mathrm{rank}(A,B)=n$ implies
that we have also $\mathrm{rank}(A-\kappa B)=n$.

Let us now concentrate on the class of couplings for which we
established in the previous section the boundary condition
convergence. In this case $A-\kappa B$ equals
 $$
\left(
\begin{array}{cccc}
    \alpha_1(\beta_1+\kappa)-1 & -(\alpha_2(\beta_2+\kappa)-1) & \cdots & 0 \\
    \vdots & & \ddots & \\
    \alpha_1(\beta_1+\kappa)-1 & 0 & \cdots & -(\alpha_n(\beta_n+\kappa)-1) \\
  (\beta_1+\kappa)\left(\frac{\omega}{n}\alpha_1-1\right)-\frac{\omega}{n} &
  (\beta_2+\kappa)\left(\frac{\omega}{n}\alpha_2-1\right)-\frac{\omega}{n} &
  \cdots & (\beta_n+\kappa)\left(\frac{\omega}{n}\alpha_n-1\right)-\frac{\omega}{n}
\end{array}
\right),
 $$
and a tedious by straightforward computation yields an explicit
form of the matrix $-(A-\kappa B)^{-1}B$, namely
\begin{equation*}
\begin{split}
\left[-(A-\kappa B)^{-1}B\right]_{jl}=&\frac{1}{\omega
-\sum^{n}_{m=1}\frac{\beta_m+\kappa}{\alpha_m(\beta_m+\kappa)-1}}
\cdot\frac{1}{(\alpha_j(\beta_j+\kappa)-1)(\alpha_l(\beta_l+\kappa)-1)} \\
& \hspace{7cm}\text{ for } j\neq l\,,\\
\left[-(A-\kappa
B)^{-1}B\right]_{jj}=&\frac{1}{\omega
-\sum^{n}_{m=1}\frac{\beta_m+\kappa}{\alpha_m(\beta_m+\kappa)-1}}
\cdot\frac{1}{(\alpha_j(\beta_j+\kappa)-1)^2}
+\frac{\alpha_j}{\alpha_j(\beta_j+\kappa)-1}\,.
\end{split}
\end{equation*}
In this way we get the Green function $\GG^{\omega,\vec{\alpha},
\vec{\beta}}_{\i\kappa}$. As we have mentioned above, it is an
$n\times n$ matrix-valued function the ($j,l$)-th element of which
is given by
\begin{equation*}
\begin{split}
\GG^{\omega,\vec{\alpha},\vec{\beta}}_{\i\kappa,jl}(x,y)=&
\delta_{jl}\left(\frac{\sinh\kappa x_<\: \e^{-\kappa x_>}}{\kappa}
+\e^{-\kappa(x+y)}\frac{\alpha_j}{\alpha_j(\beta_j+\kappa)-1}\right)+\\
+&\frac{1}{\omega-\sum^{n}_{m=1}\frac{\beta_m+\kappa}
{\alpha_m(\beta_m+\kappa)-1}}\cdot\frac{1}{(\alpha_j(\beta_j
+\kappa)-1)(\alpha_l(\beta_l+\kappa)-1)}\ \e^{-\kappa
x}\e^{-\kappa y}\,;
\end{split}
\end{equation*}
we use the convention that $x$ is from the $j$-th halfline and $y$
from the $l$-th one.

Next we will pass to resolvent construction for the approximating
family of operators $H^{u,\vec{v},\vec{w}}(d)$. As a starting
point we consider $n$ independent halflines with Dirichlet
endpoints; we know that the appropriate Green's function is
$\GG_{\i\kappa}(x,y)=\kappa^{-1} \sinh\kappa x_< \e^{-\kappa
x_>}$. The sought resolvent kernel will be then found in several
steps. Each of them represents an application of Krein's formula.
First we add the $\delta$ interaction with the parameter $v$ at
the distance $d^3$ from the endpoint, then another one with the
parameter $w$ at the distance $d+d^3$, again from the endpoint.
This is done on each halfline separately. In the final step we
find Green's function for the star in which the Dirichlet ends are
replaced by the $\delta$ coupling with the parameter $u$. That
will require, of course, to distinguish the halflines by their
indices.

The first step is rather standard \cite{CE04} and resulting Green
function is
\begin{equation}\label{aproximujici1}
\GG_{\i\kappa}^v(x,y)=\GG_{\i\kappa}(x,y)
-\frac{v}{1+v\cdot\GG_{\i\kappa}(d^3,d^3)}\ \GG_{\i\kappa}(y,d^3)\
\GG_{\i\kappa}(x,d^3)\,.
\end{equation}
Adding another $\delta$ interaction at the distance $d$ from the
previous one we seek the kernel in the form $R^{v,w}(k^2)
=R^v(k^2)+\lambda(k^2)\big(\phi(\overline{k^2}), \cdot\big)
\phi(k^2)$ where the first term is $R^v(k^2):=\GG_{\i\kappa}^v$
and the deficiency-subspace element $\phi(k^2)$ is chosen as
$$
\phi(k^2)(x):=\GG^v_{\i\kappa}(x,d+d^3)\,.
$$
We apply this Ansatz to any $\psi\in L^2(\R^+)$ and denote
$h:=R^{v,w}(k^2)\psi$. It is easy to check that $\overline{
\GG^v_{\i\bar\kappa}(x,y)}= \GG^v_{\i\kappa}(x,y)$, hence we can
write $h$ explicitly as
$$
h(x)=\int^{+\infty}_{0}\GG^v_{\i\kappa}(x,y)\psi(y)\,\D y
+\lambda(k^2)\int^{+\infty}_{0} \GG^v_{\i\kappa}(y,d+d^3)\psi(y)\,
\D y\cdot\GG^v_{\i\kappa}(x,d+d^3)\,.
$$
By definition this function this function belongs to the domain of
the operator with two $\delta$ interactions, in particular, it has
to satisfy the boundary conditions
\begin{gather}
h({d+d^3}_+)=h({d+d^3}_-)=:h(d+d^3)\,,\label{KrOP3} \\
h'({d+d^3}_+)-h'({d+d^3}_-)=w\cdot h(d+d^3)\,.\label{KrOP4}
\end{gather}
Green's function continuity implies~\eqref{KrOP3}. Furthermore, we
have
$$
h'(x)=\int^{+\infty}_{0}\frac{\partial\GG^v_{\i\kappa}(x,y)}{\partial
x}\psi(y)\, \D y
+\lambda(k^2)\int^{+\infty}_{0}\GG^v_{\i\kappa}(y,d+d^3)\psi(y)\,
\D y\cdot\frac{\partial\GG^v_{\i\kappa}(x,d+d^3)}{\partial x}\,,
$$
which allows us to express $h'(d+d^3_+)-h'(d+d^3_-)$. The first
term obviously does not contribute to the difference, while the
contribution of the second one simplifies in view of
$\left.\frac{\partial\mathcal{G}(x,y)}{\partial x}\right|_{y_+} -
\left.\frac{\partial\mathcal{G}(x,y)}{\partial x}\right|_{y_-}
=-1$ to the form
\begin{equation*}
h'(d+d^3_+)-h'(d+d^3_-)=
-\lambda(k^2)\int^{+\infty}_{0}\GG^v_{\i\kappa}(y,d+d^3)\psi(y) \,
\D y\,.
\end{equation*}
To satisfy~\eqref{KrOP4} the coefficient $\lambda(k^2)$ must obey
the condition
$$
\int^{+\infty}_{0}\left[\lambda(k^2) +w
+w\lambda(k^2)\GG^v_{\i\kappa}(d+d^3,d+d^3)\right]
\GG^v_{\i\kappa}(y,d+d^3)\psi(y)\,\D y=0
$$
for any $\psi\in L^2(\R^+)$, where we have taken Green's function
symmetry with respect to the argument interchange into account.
Consequently, the square bracket has to vanish and we get the
formula for the kernel with two $\delta$ interactions,
\begin{equation}\label{aproximujici2}
\GG_{\i\kappa}^{v,w}(x,y) =\GG^v_{\i\kappa}(x,y)
-\frac{w}{1+w\cdot\GG^v_{\i\kappa}(d+d^3,d+d^3)}\
\GG^v_{\i\kappa}(y,d+d^3)\ \GG^v_{\i\kappa}(x,d+d^3)\,.
\end{equation}

The remaining step will be more complicated because we are going
to introduce a coupling between different halflines working this
with matrix-valued functions. Our tool will be again Krein's
formula which now takes the form
$$
R_{H^{u,\vec{v},\vec{w}}}(k^2) =R_{H^{\vec{v},\vec{w}}}(k^2)
+\sum^{n}_{j,l=1}\lambda_{jl}(k^2)
\left(\phi_l(\overline{k^2}),\cdot\right)_{
L^2((\mathbb{R}^+)^n)}\cdot\phi_j(k^2)\,,
$$
where the functions $\phi_j(k^2)$ will be chosen as
$$
\left(\phi_j(k^2)(x)\right)_m
=\delta_{jm}\cdot\left.\frac{\partial\GG^{v_m,w_m}_{\i\kappa}(x,y)}
{\partial y}\right|_{y=0}\,.
$$
We apply this Ansatz to an arbitrary $\Psi=\{\psi_1,\ldots,
\psi_n\}^T$ and denote the elements of the resulting vector as
$h_j$, explicitly
\begin{equation}\label{KrForRad}
\begin{split}
h_j(x)=&\int^{+\infty}_{0}\GG^{v_j,w_j}_{\i\kappa}(x,y)
\psi_j(y)\D y\ +\\
+&\sum^{n}_{l=1}\lambda_{jl}(k^2)\int^{+\infty}_{0}
\left.\frac{\partial\GG^{v_l,w_l}_{\i\kappa}(x,y)}{\partial
x}\right|_{x=0}\psi_l(y) \,\D y
\cdot\left.\frac{\partial\GG^{v_j,w_j}_{\i\kappa}(x,y)} {\partial
y} \right|_{y=0}\,.
\end{split}
\end{equation}
where we have used Green's function symmetry and the fact that its
complex conjugation is equivalent to switching from $\kappa$ to
$\bar\kappa$. As before the functions $h_1,h_2,\ldots,h_n$ have to
satisfy the boundary conditions expressing the $\delta$ coupling
in the star centre,
\begin{gather}
h_1(0)=h_2(0)=\cdots=h_n(0)=:h(0)\,, \label{KrOP5}\\
h_1'(0)+h_2'(0)+\cdots+h_n'(0)=u\cdot h(0)\,, \label{KrOP6}
\end{gather}
for any $\psi_1,\ldots,\psi_n\in L^2(\R^+)$. Let us first express
$h_j(0)$. The first term in the above expression does not
contribute since $\GG^{v_j,w_j}_{\i\kappa}(0,y)=
\GG_{\i\kappa}^v(0,y)=0$. The second one contains the value of
Green's function derivative which can be expressed
using~\eqref{aproximujici2},
\begin{align*}
\left.\frac{\partial\GG^{v_j,w_j}_{\i\kappa}(x,y)}{\partial
y}\right|_{y=0} &=
\left.\frac{\partial\GG^v_{\i\kappa}(x,y)}{\partial y}\right|_{y=0}\\
-&\frac{w}{1+w\cdot\GG^v_{\i\kappa}(d+d^3,d+d^3)}\cdot
\left.\frac{\partial\GG^v_{\i\kappa}(y,d+d^3)}{\partial
y}\right|_{y=0}\cdot \GG^v_{\i\kappa}(x,d+d^3)\,.
\end{align*}
The first term is obtained from~\eqref{aproximujici1} together
with the explicit form of the ``free'' kernel
$\GG_{\i\kappa}(x,y)$: we have
$$
\left.\frac{\partial\GG^v_{\i\kappa}(x,y)} {\partial y}
\right|_{y=0}= \e^{-\kappa x}
-\frac{v}{1+v\cdot\GG_{\i\kappa}(d^3,d^3)}\ \e^{-\kappa d^3}\
\cdot\GG_{\i\kappa}(x,d^3)\,,
$$
in particular, $\left.\frac{\partial\GG^v_{\i\kappa}(x,y)}
{\partial y}\right|_{x=y=0}=1$. This further implies
\begin{multline}\label{G'x0}
\left.\frac{\partial\GG^{v_j,w_j}_{\i\kappa}(x,y)} {\partial
y}\right|_{y=0}= \e^{-\kappa x}-\frac{v_j}
{1+v_j\cdot\GG_{\i\kappa}(d^3,d^3)}\ \e^{-\kappa d^3}\ \GG_{\i\kappa}(x,d^3)\\
-\frac{w_j}{1+w_j\cdot\GG^{v_j}_{\i\kappa}(d+d^3,d+d^3)}
\cdot\Big(\e^{-\kappa(d+d^3)}
\\ -\frac{v_j}{1+v_j\cdot\GG_{\i\kappa}(d^3,d^3)}\ \e^{-\kappa d^3}\
\GG_{\i\kappa}(d+d^3,d^3)\Big) \cdot
\GG^{v_j}_{\i\kappa}(x,d+d^3)\,,
\end{multline}
in particular, $\left.\frac{\partial
\GG^{v_j,w_j}_{\i\kappa}(x,y)} {\partial y}\right|_{x=y=0}=1$.
Putting these results together, we can simplify the expression for
the boundary values $h_j(0)$ as follows,
$$
h_j(0)=\sum^{n}_{l=1}\lambda_{jl}(k^2)\int^{+\infty}_{0}
\left.\frac{\partial\GG^{v_l,w_l}_{\i\kappa}(x,y)}{\partial
x}\right|_{x=0}\psi_l(y)\, \D y\,.
$$
Now we can find what is required to fulfill the
conditions~\eqref{KrOP5}, i.e. $h_j(0)=h_m(0)$ for all
$j,m\in\hat{n}$. This is true provided
$$
\sum^{n}_{l=1}\left(\lambda_{jl}(k^2)
-\lambda_{ml}(k^2)\right)\int^{+\infty}_{0}
\left.\frac{\partial\GG^{v_l,w_l}_{\i\kappa}(x,y)}{\partial
x}\right|_{x=0}\psi_l(y) \,\D y=0\,,
$$
holds for any $n$-tuple of functions $\psi_1,\ldots,\psi_n\in
L^2(\R^+)$ which is possible if
\begin{equation*}
\lambda_{jl}(k^2)=\lambda_{ml}(k^2) \quad \text{ for all }
j,m\in\hat{n},\; l\in\hat{n}\,,
\end{equation*}
thus we can simplify notation writing
$\lambda_l:=\lambda_{jl}(k^2)$ for a fixed $l\in\hat{n}$.

Values of the coefficients $\lambda_1,\ldots,\lambda_n$ can be
found from the remaining condition~\eqref{KrOP6}. To this aim we
have to find explicit form of $h_j'(0)$. It follows from the
expression~\eqref{KrForRad} for $h_j(x)$ that
\begin{equation*}
\begin{split}
h_j'(0)=&\int^{+\infty}_{0}
\left.\frac{\partial\GG^{v_j,w_j}_{\i\kappa}(x,y)}
{\partial x}\right|_{x=0}\psi_j(y)\,\D y\ \\
+&\sum^{n}_{l=1}\lambda_l\int^{+\infty}_{0}
\left.\frac{\partial\GG^{v_l,w_l}_{\i\kappa}(x,y)}{\partial
x}\right|_{x=0}\psi_l(y)\,\D y\cdot\left.\frac{\D}{\D
x}\left(\left.\frac{\partial\GG^{v_j,w_j}_{\i\kappa}(x,y)}
{\partial y}\right|_{y=0}\right)\right|_{x=0}\,,
\end{split}
\end{equation*}
The boundary condition~\eqref{KrOP6} then requires that the
expression
\begin{equation*}
\begin{split}
\sum^{n}_{l=1}\int^{+\infty}_{0}
\left(1+\lambda_l\sum^{n}_{j=1}\left.\frac{\D}{\D
x}\left(\left.\frac{\partial\GG^{v_j,w_j}_{\i\kappa}(x,y)}
{\partial y}\right|_{y=0}\right)\right|_{x=0}
-u\cdot\lambda_l\right) \\ \cdot \left.
\frac{\partial\GG^{v_l,w_l}_{\i\kappa}(x,y)} {\partial
x}\right|_{x=0}\psi_l(y) \,\D y
\end{split}
\end{equation*}
vanishes for any $\psi_1,\ldots,\psi_n$, and this in turn yields
$$
\lambda_l=\left[u-\sum^{n}_{j=1}\left.\frac{\D}{\D
x}\left(\left.\frac{\partial\GG^{v_j,w_j}_{\i\kappa}(x,y)}{\partial
y}\right|_{y=0}\right)\right|_{x=0} \right]^{-1} \quad \text{ for
all } l\in\hat{n}
$$
showing, in particular, that $\lambda_l$ does not depend on $l$,
which means that all the coefficients $\lambda_{jl}(k^2)$ are the
same and equal to the right-hand side of the last relation.

Before specifying the expression in the square bracket let us
write down the formula for the ($j,l$)-th component of the sought
Green function: we have

\begin{equation}\label{Guvw}
\GG^{u,\vec{v},\vec{w}}_{\i\kappa,jl}(x,y)=
\delta_{jl}\cdot\GG^{v_j,w_j}_{\i\kappa}(x,y)
+\frac{\left.\frac{\partial\GG^{v_j,w_j}_{\i\kappa}(x,y)}
{\partial y}\right|_{y=0}
\cdot\left.\frac{\partial\GG^{v_l,w_l}_{\i\kappa}(x,y)} {\partial
x} \right|_{x=0}}{u-\sum^{n}_{m=1}\left.\frac{\D}{\D x}
\left(\left.\frac{\partial\GG^{v_m,w_m}_{\i\kappa}(x,y)}{\partial
y} \right|_{y=0}\right)\right|_{x=0}}\,.
\end{equation}
The first derivative in the numerator was found in~\eqref{G'x0}
and by Green's function symmetry the other one is given by the
same expression, with $y$ replaced by $x$. The same relation
allows us to compute $\frac{\D}{\D x} \left(\left.\frac{
\partial\GG^{v_m,w_m}_{\i\kappa}(x,y)} {\partial
y}\right|_{y=0}\right)$, in particular, to evaluate the quantity
appearing in the square bracket above,
\begin{multline}\label{G''00}
\left.\frac{\D}{\D x}\left( \left.\frac{
\partial\GG^{v_m,w_m}_{\i\kappa}(x,y)}{\partial
y}\right|_{y=0}\right) \right|_{x=0}=
-\kappa-\frac{v_m}{1+v_m\cdot\GG_{\i\kappa}(d^3,d^3)}\
\e^{-\kappa d^3}\ \cdot\e^{-\kappa d^3} \\
-\frac{w_m}{1+w_m\cdot\GG^{v_m}_{\i\kappa}(d+d^3,d+d^3)}
\cdot\Big(\e^{-\kappa (d+d^3)} \\ -\frac{v_m}
{1+v_m\cdot\GG_{\i\kappa}(d^3,d^3)}\ \e^{-\kappa d^3}\
\GG_{\i\kappa}(d+d^3,d^3)\Big)^2\,.
\end{multline}
The relations~\eqref{Guvw} and~\eqref{G''00} together
with~\eqref{G'x0} and its mirror counterpart describe completely
Green's function $\GG^{u,\vec{v},\vec{w}}_{\i\kappa}$ of the
approximating operators.

After deriving explicit expressions for the resolvent we can pass
to our proper goal which is to prove that the matrix-valued kernel
$\GG^{u,\vec{v},\vec{w}}_{\i\kappa}$ converges to
$\GG^{\omega,\vec{\alpha},\vec{\beta}}_{\i\kappa}$ as $d\to 0_+$
which in terms of their components can be written as
$$
\lim_{d\to0+}\left\|\GG^{u,\vec{v},\vec{w}}_{\i\kappa,jl}
-\GG^{\omega,\vec{\alpha},\vec{\beta}}_{\i\kappa,jl}
\right\|_{L^2(\R^+\times\R^+)}=0\,.
$$
Depending on the values $x,\,y$ the difference
$\GG^{u,\vec{v},\vec{w}}_{\i\kappa,jl}(x,y) -\GG^{\omega,
\vec{\alpha},\vec{\beta}}_{\i\kappa,jl}(x,y)$ takes different
forms. Notice that one can suppose without loss of generality that
$x\leq y$, and therefore there are six different situations to
inspect, namely
\begin{itemize}
\item $\;d+d^3\leq x\leq y$, \item $\;d\leq x\leq d+d^3\leq y$,
\item $\;0<x\leq d^3$, $d+d^3\leq y$, \item $\;d^3\leq x\leq y\leq
d+d^3$, \item $\;0<x\leq d^3\leq y\leq d+d^3$, \item $\;0<x\leq
y\leq d^3$.
\end{itemize}
To express the kernel difference we employ Taylor expansion of
$\GG^{u,\vec{v},\vec{w}}_{\i\kappa,jl}(x,y)$. Let us start with
expressions which appear in the formulae repeatedly. The first one
is
$$
\frac{v_m}{1+v_m\cdot\GG_{\i\kappa}(d^3,d^3)}=
\frac{-\frac{1}{d^3}+\frac{\alpha_m}{d^2}}
{1+\left(-\frac{1}{d^3}+\frac{\alpha_m}{d^2}\right)\cdot\frac{\sinh\kappa
d^3 \:\e^{-\kappa d^3}}{\kappa}}=(*)
$$
Using $\sinh(x)=x+\OO(x^2)$ and $\e^{x}=1+\OO(x)$ we get
$$
\frac{\sinh\kappa d^3\: \e^{-\kappa d^3}}{\kappa}=\frac{(\kappa
d^3+\OO(d^6))(1+\OO(d^3))}{\kappa}=d^3(1+\OO(d^3))\,,
$$
and this in turn allows us to express (*) as follows,
\begin{equation*}
(*)=-\frac{1}{d^3}\cdot\frac{1 -\alpha_m d}
{1+\left(-\frac{1}{d^3}+\frac{\alpha_m}{d^2}\right)
\cdot(d^3(1+\OO(d^3)))}=-\frac{1}{d^4}
\cdot\left(\frac{1}{\alpha_m}+\OO(d)\right)\,.
\end{equation*}
The next frequent expression is $w_m \left(1+w_m\cdot
\GG^{v_m}_{\i\kappa}(d+d^3,d+d^3)\right)^{-1}$. We employ relation
(\ref{aproximujici1}) with $v=v_m$ and the expansion
$\e^x=1+x+\OO(x^2)$ together with the explicit form of
$\GG_{\i\kappa}$; this yields after a straightforward computation
$$
\GG^{v_m}_{\i\kappa}(d+d^3,d+d^3)= d\left(1-\kappa
d-\frac{d}{\alpha_m}+\OO(d^2)\right)= \GG_{\i\kappa}(d+d^3,d+d^3)
-\frac{d^2}{\alpha_m} +\OO(d^3)\,,
$$
and therefore
$$
\frac{w_m}{1+w_m\cdot\GG^{v_m}_{\i\kappa}(d+d^3,d+d^3)}=
-\frac{1}{d^2}\left(\frac{1}{\beta_m+\kappa-\frac{1}
{\alpha_m}}+\OO(d)\right)\,.
$$
Now we can expand the first term in $\GG^{u,\vec{v},
\vec{w}}_{\i\kappa,jl}(x,y)$. Using (\ref{aproximujici2}) for the
parameters $v=v_j,\, w=w_j$ together with the previous result we
get
\begin{equation*}
\begin{split}
&\GG_{\i\kappa}^{v_j,w_j}(x,y)= \frac{\sinh\kappa x\:\e^{-\kappa
y}}{\kappa} \\ & +\frac{1}{d^2}\left(\frac{1}{\beta_m+\kappa
-\frac{1}{\alpha_m}}+\OO(d)\right)\
\frac{\sinh\kappa(d+d^3)\,\e^{-\kappa y}}{\kappa}\
\frac{\sinh\kappa(d+d^3)\,\e^{-\kappa x}}{\kappa}
\\ & =\frac{\sinh\kappa x\:\e^{-\kappa y}} {\kappa}+\frac{1}
{\beta_m+\kappa-\frac{1}{\alpha_m}}\ \e^{-\kappa x}\ \e^{-\kappa
y}\: \big(1+\OO(d) \big)
\end{split}
\end{equation*}
As for the second term in (\eqref{Guvw}), we first expand the
derivative in the denominator using  $\GG_{\i\kappa}(d+d^3,d^3)=
d^3(1+\OO(d))$ and (\ref{G''00}). A direct computation yields
$$
\left.\frac{\D}{\D x}\left(\left.\frac{
\partial\GG^{v_m,w_m}_{\i\kappa}(x,y)}{\partial
y}\right|_{y=0}\right)\right|_{x=0}=
\frac{1}{d^4}\cdot\left(\frac{\beta_m+\kappa}
{\alpha_m(\beta_m+\kappa)-1}+\OO(d)\right)\,,
$$
and therefore
$$
\left(u-\sum^{n}_{m=1}\left.\frac{\D}{\D
x}\left(\left.\frac{\partial\GG^{v_m,w_m}_{\i\kappa}(x,y)}{\partial
y}\right|_{y=0}\right)\right|_{x=0} \right)^{-1} =
d^4\left(\frac{1}{\omega-\sum^{n}_{m=1}
\frac{\beta_m+\kappa}{\alpha_m(\beta_m+\kappa)-1}}+\OO(d)\right)\,.
$$
Next we expand the derivatives which appear in the numerator using
the relation $\GG_{\i\kappa}^{v_m}(x,d+d^3) =d(1+\OO(d))\,
\e^{-\kappa x}$; it gives
\begin{multline*}
\left.\frac{\partial\GG^{v_m,w_m}_{\i\kappa}(x,y)} {\partial
y}\right|_{y=0} =
\e^{-\kappa x}-\frac{v_m}{1+v_m\cdot\GG_{\i\kappa}(d^3,d^3)}\
\e^{-\kappa d^3}\ \GG_{\i\kappa}(x,d^3)-\\
-\frac{w_m}{1+w_m\cdot\GG^{v_m}_{\i\kappa}(d+d^3,d+d^3)}\ \cdot\\
=\frac{1}{d^2}\left(\frac{1}{\alpha_m(\beta_m+\kappa)-1}
+\OO(d)\right)\e^{-\kappa x}\,.
\end{multline*}
and the analogous expression for $\left.\frac{\partial
\GG^{v_m,w_m}_{\i\kappa}(x,y)}{\partial x}\right|_{x=0}$ with $x$
replaced by $y$. This determines the behaviour of the second term
at the right-hand side of (\eqref{Guvw}) as $d\to 0_+$, and for
the full kernel $\GG^{u,\vec{v}, \vec{w}}_{\i\kappa,jl}(x,y)$ we
consequently have
\begin{multline*}
\begin{split}
& \GG^{u,\vec{v},\vec{w}}_{\i\kappa,jl}(x,y)=
\delta_{jl}\left(\frac{\sinh\kappa x\ \e^{-\kappa
y}}{\kappa}+\frac{1+\OO(d)}{\beta_j+\kappa-\frac{1}{\alpha_j}}\:
\e^{-\kappa x}\ \e^{-\kappa y} \right)\ \\
& +\left(\frac{1}{\omega-\sum^{n}_{m=1}\frac{\beta_m+\kappa}
{\alpha_m(\beta_m+\kappa)-1}}\cdot
\frac{1}{\alpha_j(\beta_j+\kappa)-1}\cdot
\frac{1}{\alpha_l(\beta_l+\kappa)-1}+\OO(d)\right)\e^{-\kappa
x}\e^{-\kappa y}\,.
\end{split}
\end{multline*}
On the other hand, for $x\leq y$ we have
\begin{equation*}
\begin{split}
\GG^{\omega,\vec{\alpha},\vec{\beta}}_{\i\kappa,jl}(x,y)=&
\delta_{jl}\left(\frac{\sinh\kappa x\ \e^{-\kappa y}}{\kappa}
+\e^{-\kappa(x+y)}\frac{\alpha_j}{\alpha_j(\beta_j+\kappa)-1}\right)+\\
+&\frac{1}{\omega-\sum^{n}_{m=1}\frac{\beta_m+\kappa}
{\alpha_m(\beta_m+\kappa)-1}}\cdot\frac{1}
{(\alpha_j(\beta_j+\kappa)-1)(\alpha_l(\beta_l+\kappa)-1)}\
\e^{-\kappa x}\e^{-\kappa y}\,,
\end{split}
\end{equation*}
hence the Green function difference satisfies
\begin{equation*}
\GG^{u,\vec{v},\vec{w}}_{\i\kappa,jl}(x,y)
-\GG^{\omega,\vec{\alpha},\vec{\beta}}_{\i\kappa,jl}(x,y)
=\OO(d)\: \e^{-\kappa x}\e^{-\kappa y} \quad \textrm{as }\: d\to
0_+\,.
\end{equation*}
The same estimate is obviously valid also for $d<y<x$, hence there
is a constant $K$ independent of $d,\,x$ and $y$ such that
\begin{equation}\label{d+d3<x<y}
\left|\GG^{u,\vec{v},\vec{w}}_{\i\kappa,jl}(x,y)
-\GG^{\omega,\vec{\alpha},\vec{\beta}}_{\i\kappa,jl}(x,y)\right|
<K\,d\:\e^{-\kappa x}\e^{-\kappa y}
\end{equation}
holds for all $d<1,\: x\ge d+d^3$ and $y\ge d+d^3$. Now we are in
position to estimate the Hilbert-Schmidt norm of the resolvent
difference for the operators $H^{\omega,\vec{\alpha},\vec{\beta}}$
and $H^{u,\vec{v},\vec{w}}(d)$ which can be written explicitly as
follows,
\begin{equation*}
\begin{split}
\left\|R_{H^{u,\vec{v},\vec{w}}(d)}(k^2)
-\right.&\left.R_{H^{\omega,\vec{\alpha},\vec{\beta}}}\right\|_2^2=
\sum^{n}_{j,l=1}\int^{+\infty}_{0}\int^{+\infty}_{0}\left|
\GG^{u,\vec{v},\vec{w}}_{\i\kappa,jl}(x,y)
-\GG^{\omega,\vec{\alpha},\vec{\beta}}_{\i\kappa,jl}(x,y)\right|^2
\,\D x\D y\\
=\sum^{n}_{j,l=1}&\left(\int^{+\infty}_{d+d^3}\int^{+\infty}_{d+d^3}
\left|\GG^{u,\vec{v},\vec{w}}_{\i\kappa,jl}(x,y)
-\GG^{\omega,\vec{\alpha},\vec{\beta}}_{\i\kappa,jl}(x,y)\right|^2
\,\D x\D y\right.\\
&+\int^{d+d^3}_{d^3}\int^{+\infty}_{d+d^3}\left|
\GG^{u,\vec{v},\vec{w}}_{\i\kappa,jl}(x,y)
-\GG^{\omega,\vec{\alpha},\vec{\beta}}_{\i\kappa,jl}(x,y)\right|^2
\,\D x\D y\\
&+\int^{+\infty}_{d+d^3}\int^{d+d^3}_{d^3}\left|
\GG^{u,\vec{v},\vec{w}}_{\i\kappa,jl}(x,y)
-\GG^{\omega,\vec{\alpha},\vec{\beta}}_{\i\kappa,jl}(x,y)\right|^2
\,\D x\D y\\
&+\int^{d^3}_{0}\int^{+\infty}_{d+d^3}\left|
\GG^{u,\vec{v},\vec{w}}_{\i\kappa,jl}(x,y)
-\GG^{\omega,\vec{\alpha},\vec{\beta}}_{\i\kappa,jl}(x,y)\right|^2
\,\D x\D y\\
&+\int^{+\infty}_{d+d^3}\int^{d^3}_{0}\left|
\GG^{u,\vec{v},\vec{w}}_{\i\kappa,jl}(x,y)
-\GG^{\omega,\vec{\alpha},\vec{\beta}}_{\i\kappa,jl}(x,y)\right|^2
\,\D x\D y\\
&+\int^{d+d^3}_{d^3}\int^{d+d^3}_{d^3}\left|
\GG^{u,\vec{v},\vec{w}}_{\i\kappa,jl}(x,y)
-\GG^{\omega,\vec{\alpha},\vec{\beta}}_{\i\kappa,jl}(x,y)\right|^2
\,\D x\D y\\
&+\int^{d^3}_{0}\int^{d+d^3}_{d^3}\left|
\GG^{u,\vec{v},\vec{w}}_{\i\kappa,jl}(x,y)
-\GG^{\omega,\vec{\alpha},\vec{\beta}}_{\i\kappa,jl}(x,y)\right|^2
\,\D x\D y\\
&+\int^{d+d^3}_{d^3}\int^{d^3}_{0}\left|
\GG^{u,\vec{v},\vec{w}}_{\i\kappa,jl}(x,y)
-\GG^{\omega,\vec{\alpha},\vec{\beta}}_{\i\kappa,jl}(x,y)\right|^2
\,\D x\D y\\
&\left.+\int^{d^3}_{0}\int^{d^3}_{0}\left|
\GG^{u,\vec{v},\vec{w}}_{\i\kappa,jl}(x,y)
-\GG^{\omega,\vec{\alpha},\vec{\beta}}_{\i\kappa,jl}(x,y)\right|^2
\,\D x\D y\right)\,.
\end{split}
\end{equation*}
The inequality~\eqref{d+d3<x<y} makes it possible to estimate the
first one of the integrals,
\begin{multline*}
\int^{+\infty}_{d+d^3}\int^{+\infty}_{d+d^3}\left|
\GG^{u,\vec{v},\vec{w}}_{\i\kappa,jl}(x,y)
-\GG^{\omega,\vec{\alpha},\vec{\beta}}_{\i\kappa,jl}(x,y)\right|^2
\,\D x\D y\leq
K^2d^2\:\left(\int^{+\infty}_{d+d^3}e^{-2(\Re\kappa)x}\,\D x\right)^2\\
=K^2d^2\frac{e^{-2(\Re\kappa)(d+d^3)}}{2\Re\kappa}
\leq\frac{K^2}{2\Re\kappa}\:d^2\,,
\end{multline*}
and it is obvious from this inequality that for $d\to0_+$ the
integral tends to zero for any $j,l\in\hat{n}$. In a similar way
one can estimate each of the remaining eight integrals: using
Taylor expansions of $\GG^{u,\vec{v},\vec{w}}_{\i\kappa,jl}$ we
get a bound for the integrand which shows that the integral
vanishes as $d\to0_+$. Since the argument repeats the procedure
described above, we skip the details. Putting all this together,
we conclude that
$$
\lim_{d\to 0_+}
\left\|R_{H^{u,\vec{v},\vec{w}}(d)}(k^2)
-R_{H^{\omega,\vec{\alpha},\vec{\beta}}}\right\|_2^2=0\,,
$$
and therefore the resolvent difference tends to zero in
Hilbert-Schmidt norm as $d\to 0_+$ which is what we set up to
demonstrate.
\end{proof}

\section{Approximations with added edges}

We have seen that a CS--type scheme can produce a $2n$-parameter
family of (self-adjoint) couplings out of the whole set depending
on $n^2$ real numbers. To get a wider class we have to add to the
star graph $\Gamma$ not only vertices but edges as well.

\subsection{Admissible couplings}

The first question naturally is how many parameters can be
achieved in this way. An upper bound on this number is given by
the following statement.

\begin{proposition} \label{edge-admissible}
Let $\Gamma$ be a star graph with $n$ semi-infinite edges and
denote by $\left\{\tilde\Gamma(d):\: d\in\R^+ \right\}$ a family
of graphs obtained from $\Gamma$ by adding finite edges connecting
pairwise the halflines; their number may be arbitrary finite but
independent of $d$. Suppose that $\tilde\Gamma(d)$ supports only
$\delta$ couplings and $\delta$ interactions, their number again
independent of $d$, and that the distances between all their sites
are $\OO(d)$ as $d\to 0_+$. Suppose that a family of functions
$\Psi_d \in W^{2,2}\left(\Gamma \setminus(\{c\}\cup V_d)\right)$,
where $c$ is the centre of $\Gamma$, and $V_d$ is the set of the
vertices added on the halflines, satisfies the conditions
(\ref{delta}) with $d$-dependent parameters, and that it converges
to $\Psi\in W^{2,2}\left(\Gamma \setminus\{c\}\right)$ which obeys
the condition (\ref{1}) with some $A,\,B$ satisfying the
requirements (\ref{OP}). The family of the conditions (\ref{1})
which can be obtained in this way has real-valued coefficients,
$A,B\in\R^{n,n}$, depending thus on at most ${n+1\choose 2}$
parameters.
\end{proposition}

\begin{proof}
The $\delta$ coupling in the centre of $\tilde\Gamma(d)$,
identified with centre of $\Gamma$, is expressed by the conditions
(\ref{delta}). For any $j\in\hat{n}$ we denote by $d_j$ the
coordinate of the most distant point on the $j$-th halfline which
supports either a $\delta$ interaction or a $\delta$ coupling at
the endpoint of an added edge. We arrange the function values at
these points into the $n$-tuple $\Psi(d)$, and similarly
$\Psi'(d_+)$ is the $n$-tuple of right derivatives. Let us stress
that this a symbolic notation; the elements are $\psi_j(d_j)$ and
$\psi'_j(d_{j+})$, respectively.

As in the proof of Proposition~\ref{admissible} we can use
(\ref{delta}) to express these quantities through the common value
$\psi(0)$ and the right derivatives $\Psi'(0_+)$ at the origin
\begin{eqnarray*}
M_1(d)\Psi(d) &=& \psi(0)\cdot m_2(d)+M_3(d)\Psi'(0)+\RR(d)\,,
\\ N_1(d)\Psi'(d_+) &=& \psi(0)\cdot
n_2(d)+N_3(d)\Psi'(0)+\tilde{\RR}(d) 
\end{eqnarray*}
for some $M_1,M_3,N_1,N_3:\R^+\to\R^{n,n}$,
$\:m_2,n_2:\R^+\to\R^n$ and error terms $\RR, \tilde{\RR}:
\R^+\to\R^n$ supposed to be negligible as $d\to0_+$; we may assume
that $\RR,\, \tilde{ \RR}=o(1)$. The above system can be also
written in a matrix form,
\begin{equation*}
\left(
\begin{array}{cccc}
    M_1(d) & 0 & -m_2(d) & -M_3(d) \\
    0 & N_1(d) & -n_2(d) & -N_3(d) \\
  0 & 0 & -\alpha &
\begin{array}{cccc}
    1 & 1 & \cdots & 1
\end{array}
\end{array}
\right) \left(
\begin{array}{c}
\Psi(d) \\ \Psi'(d_+) \\ \psi(0) \\ \Psi'(0)
\end{array}
\right) = \left(
\begin{array}{c}
o(1) \\ o(1) \\ 0
\end{array}
\right)
\end{equation*}
To find an approximation in the described sense one has to find a
relation between $\Psi(d)$ and $\Psi'(d_+)$ eliminating
$\psi(0),\,\Psi'(0)$. Since the former are determined by the
latter we may suppose that the matrices $M_1(d)$ a $N_1(d)$ are
regular; the elimination then leads to a system
$$
A(d)\Psi(d)+B(d)\Psi'(d_+)=\breve{\RR}(d)\,,
$$
where the matrices $A(d)$, $B(d)$ are real for all $d\in\R^+$ and
the right-hand side consists of an error term
$\breve{\RR}:\R^+\to\R^n$. We multiply the last equation by a
power of $d$ such that the right-hand side is $o(1)$ as $d\to0_+$
while the left-hand one has a nontrivial limit. It is clear that
we can get in this way the condition (\ref{parametrizace}) with
real-valued coefficients, $A,B\in\R^{n,n}$.
\end{proof}

\subsection{A concrete approximation arrangement}

The above discussion leaves open the question how such an
approximation can be constructed to cover the mentioned
${n+1\choose 2}$-parameter family. Our aim here is to demonstrate
a specific way to do that. We consider the coupling
(\ref{parametrizace}) with real $A,\,B$, and for simplicity we
restrict our attention only to the generic case assuming that $B$
is regular so that the boundary conditions acquire the form
 $$
\Psi'(0)=-B^{-1}A\Psi(0)
 $$
with a symmetric matrix $-B^{-1}A$. We can also write them as
\begin{equation}\label{3}
\Psi'(0)=(D+S)\Psi(0)\,,
\end{equation}
where the real matrix $D$ is diagonal while $S$ is real symmetric
with a vanishing diagonal; it is clear that $D$ and $S$ depend on
$n$ and ${n\choose 2}$ real parameters, respectively.

\begin{figure}[h]
\begin{center}
\includegraphics[angle=0,width=0.5\textwidth]{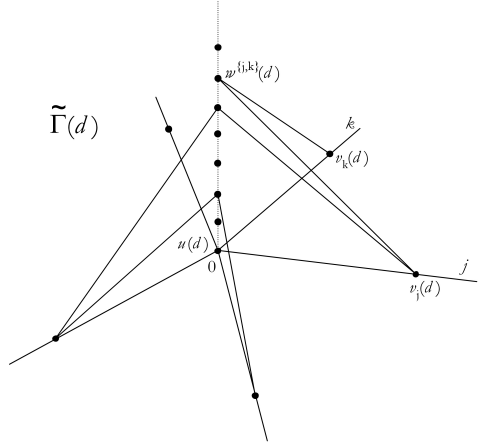}
\caption{Approximating graph: a star amended by connections of the
edges, with a $\delta$ coupling in the centre, one $\delta$
coupling at each edge and one $\delta$ interaction at each
(broken) connection segment}\label{Usporadani}
\end{center}
\end{figure}

\begin{figure}[ht]
\begin{center}
\includegraphics[angle=0,width=0.4\textwidth]{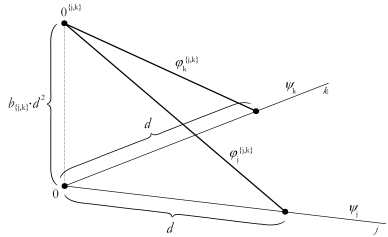}
\caption{The connecting edge between the $j$-th and $k$-th
halfline}\label{Napojeni}
\end{center}
\end{figure}

To construct approximation of the corresponding operator $H^{A,B}$
we have find suitable family of graphs $\tilde\Gamma(d)$. The
decomposition of the matrix in (\ref{3}) into the diagonal and
off-diagonal part inspires the following scheme:
\begin{itemize}
 \item the centre of $\Gamma$ supports a $\delta$ coupling with
the parameter $u(d)$ the dependence of which on $d$ will be
specified below
 \item at each edge of $\Gamma$ we place a $\delta$ coupling at
the distance $d$ from the centre; the corresponding parameter
$v_j(d)$, to be again specified, will be related to the diagonal
element $D_{jj}$ of the matrix $D$
 \item the pairs of edges whose indices $j,k$ correspond to nonzero
elements of the matrix $S$ we join by an additional edge, whose
endpoints are the $\delta$ coupling sites mentioned above, and in
the middle of this edge we place the $\delta$ interaction with a
parameter $w_{\{j,k\}}(d)$ related to the value of $S_{jk}$
\end{itemize}
The metric on $\Gamma$ and $\tilde\Gamma(d)$ is intrinsic,
nevertheless, it is useful to think of it as of induced by
embedding of the graphs into a Euclidean space. Without loss of
generality we may consider the original star $\Gamma$ as a planar
graph and to construct as embedded into $\R^3$. In such a case, of
course, we have to make sure that the added edges do not
intersect. This can be achieved in the way sketched in
Fig.~\ref{Usporadani}.
A possible way is to employ the bijection $b$ from the family of
two-element subsets of $\{1,2,\ldots n\}$ to the set $\{1,2,\ldots
\frac{n(n-1)}{2}\}$. The edge connecting the $j$-th and $k$-th
halfline is formed by two segments connected in a V-shape. Its
endpoints are at the $j$-th and $k$-th halfline, both at the
distance $d$ from the centre. The tip of this $V$-graph is placed
on the halfline starting from the centre of $\Gamma$ in the
perpendicular direction to its plane -- see Fig.~\ref{Napojeni} --
at the distance $b_{jk}\cdot d^2$, so that the length of the
connecting V-graph is $d\sqrt{1+(b_{jk}d)^2}$.

As before we denote by $\psi_j$ the wave function on the $j$-th
halfline assuming that all the coordinates have zero in the centre
of $\Gamma$. Furthermore, we denote by $\varphi^{\{j,k\}}_j$ and
$\varphi^{\{j,k\}}_k$ the wave function on the line segment part
of the connection between the $j$-th and $k$-th halfline which is
attached by one of its endpoints to the $j$-th and $k$-th
halfline, respectively; notice that the order of the upper indices
is irrelevant. Such a connecting link is regarded as a star with
two edges of the same length. For the sake of brevity we introduce
also the set $N_j$ defined as
$$
N_j=\{k\in\hat{n}:\: S_{jk}\neq 0\}\,;
$$
its cardinality $\#N_j$ tells us how many nonzero elements are in
the $j$-th row of the matrix $S$, in other words, how many
V-shaped connecting edges sprout from the point $x_j=d$ on the
$j$-th halfline.

Next we will write down the boundary conditions describing the
involved $\delta$ couplings; for simplicity we will not indicate
the dependence of the parameters $u,\, v_j,\, w_{\{j,k\}}$ on the
distance $d$. The $\delta$ coupling in the centre of $\Gamma$
means
\begin{equation}\label{I.serie}
\psi_1(0)=\psi_2(0)=\cdots=\psi_n(0)=:\psi(0)\,, \quad
\sum^{n}_{j=1}\psi_j'(0_+)=u\psi(0)\,,
\end{equation}
the $\delta$ interaction at the ``tip'' of the broken edge
connecting the $j$-th and $k$-th halfline between the vertices
added at the distance $d$ from the centre (of course, for
$j,k\in\hat{n}$ such that $S_{jk}\neq 0$ only) is expressed
through the conditions
\begin{equation}\label{Ia.serie}
\varphi^{\{j,k\}}_j(0)=\varphi^{\{j,k\}}_k(0)=:\varphi^{\{j,k\}}(0)\,,
\;\; (\varphi^{\{j,k\}})_j'(0_+)+(\varphi^{\{j,k\}})_k'(0_+)
=w_{\{j,k\}}\varphi^{\{j,k\}}(0)\,,
\end{equation}
and finally, the $\delta$ coupling at the mentioned added vertices
added requires
\begin{equation}\label{II.serie}
\begin{array}{c}
\psi_j(d_+)=\psi_j(d_-)
=\varphi^{\{j,k\}}_j\big(d\sqrt{1+(b_{jk}d)^2}\big) =:\psi_j(d)\,,
\quad j\in\hat{n}\,, k\in N_j \\ \vspace{.5em}
\psi_j'(d_+)-\psi_j'(d_-)-\sum_{k\in
N_j}(\varphi^{\{j,k\}}_j)'\big(d\sqrt{1+(b_{jk}d)^2}_-\big)
=v_j\psi_j(d)\,, \quad j\in\hat{n}\,.
\end{array}
\end{equation}

Further relations which will help us to find the parameter
dependence on $d$ come from Taylor expansion,
\begin{eqnarray}\label{III.serie}
&&\psi_j(d)=\psi_j(0)+d\psi_j'(0)+\OO)(d^2)\,, \quad
\psi_j'(d_-)=\psi_j'(0_+)+\OO(d)\,, \quad j\in\hat{n}\,, \\
&& \begin{split}
&\varphi^{\{j,k\}}_j\left(d\sqrt{1+(b_{jk}d)^2}\right)
=\varphi^{\{j,k\}}(0)+d\sqrt{1+(b_{jk}d)^2}\,
(\varphi^{\{j,k\}}_j)'(0_+)+\OO(d^2)\,,\\
&(\varphi^{\{j,k\}}_j)'\left(d\sqrt{1+(b_{jk}d)^2}_-\right)
=(\varphi^{\{j,k\}}_j)'(0_+)+\OO(d)\,, \quad j,k\in\hat{n}\,,
\label{IIIa.serie}
\end{split}
\end{eqnarray}
where we have used the fact that $\sqrt{1+(b_{jk}d)^2}
=1+\OO(d^2)$. Now we employ the first of the
relations~\eqref{IIIa.serie} together with the
continuity~\eqref{II.serie}, which yields
\begin{equation}\label{hvezdicka}
d\sqrt{1+(b_{jk}d)^2}\: (\varphi^{\{j,k\}}_j)'(0_+)=\psi_j(d)
-\varphi^{\{j,k\}}(0)+\OO(d^2)\,.
\end{equation}
The same relation holds with $j$ replaced by $k$, summing them
together and using the second of the relations~\eqref{Ia.serie} we
get
$$
\left(2+d\sqrt{1+(b_{jk}d)^2}\:
w_{\{j,k\}}\right)\varphi^{\{j,k\}}(0)
=\psi_j(d)+\psi_k(d)+\OO(d^2)\,.
$$
We express $\varphi^{\{j,k\}}(0)$ from here and substitute
into~\eqref{hvezdicka} obtaining
\begin{equation}\label{phi'0+}
d\sqrt{1+(b_{jk}d)^2}\: (\varphi^{\{j,k\}}_j)'(0_+)
=\psi_j(d)-\frac{\psi_j(d)+\psi_k(d)+\OO(d^2)}
{2+d\sqrt{1+(b_{jk}d)^2}\cdot w_{\{j,k\}}}+\OO(d^2)\,.
\end{equation}
The relations ~\eqref{III.serie} and~\eqref{I.serie} give
\begin{equation}\label{xxxx}
d\psi_j'(0_+)=\psi_j(d)-\psi(0)+\OO(d^2)\,,
\end{equation}
and summing this over
$j\in\hat{n}$ we arrive at the identity
$$
d\sum^{n}_{j=1}\psi_j'(0_+)
=\sum^{n}_{j=1}\psi_j(d)-n\psi(0)+\OO(d^2)\,.
$$
The right-hand side of it can be rewritten using~\eqref{I.serie}.
This makes it possible to express $\psi(0)$; substituting it into~\eqref{xxxx} we get
\begin{equation}\label{dpsij'(0+)}
d\psi_j'(0_+)=\psi_j(d)
-\frac{\sum^{n}_{k=1}\psi_k(d)+\OO(d^2)}{n+du}+\OO(d^2)\,.
\end{equation}
Next we use consecutively the second relations
of~\eqref{II.serie}, \eqref{III.serie} and~\eqref{IIIa.serie} to
infer
\begin{equation*}
\begin{split}
\psi_j'(d_+)=&v_j\psi_j(d)+\sum_{k\in N_j} (\varphi^{\{j,k\}})_k'\,
\Big(d\sqrt{1+(b_{jk}d)^2}_-\Big)+\psi_j'(d_-)\\
=&v_j\psi_j(d)+\sum_{k\in N_j} (\varphi^{\{j,k\}})_k'(0_+)
+\psi_j'(0_+)+\OO(d)\,.
\end{split}
\end{equation*}
Substituting into the last relation from~\eqref{phi'0+}
and~\eqref{dpsij'(0+)} we get
\begin{equation*}
\begin{split}
\psi_j'(d_+)=&\left(v_j+\frac{1}{d} \left(\sum_{k\in N_j}\frac{1}
{\sqrt{1+(b_{jk}d)^2}}+1\right)\right)\psi_j(d) \\
&-\frac{1}{d}\sum_{k\in N_j}\frac{1}{\sqrt{1+(b_{jk}d)^2}}\cdot
\frac{\psi_j(d)+\psi_k(d)}{2+d\sqrt{1+(b_{jk}d)^2}\cdot w_{\{j,k\}}} \\
&-\frac{1}{d(n+du)}\left(\sum^{n}_{k=1}\psi_k(d)+\OO(d^2)\right)+\OO(d)\,,
\end{split}
\end{equation*}
where we have also employed the fact that
$\OO(d)\,[1+(b_{jk}d)^2]^{-1/2}=\OO(d)$ holds as $d\to 0_+$ for
all $j\neq k$, $j,k\in\hat{n}$.

Now we can finally ask about the parameter dependence on $d$.
Since the last relation is supposed to yield in the limit $d\to
0_+$ the $j$-th row of the matrix condition (\ref{3}), it would be
sufficient to have the following requirements satisfied:
\begin{equation}\label{one}
\lim_{d\to0_+}\left(v_j+\frac{1}{d}\left(\sum_{k\in N_j}
\frac{1}{\sqrt{1+(b_{jk}d)^2}}+1-\sum_{k\in N_j}
\frac{1}{2+d\sqrt{1+(b_{jk}d)^2}w_{\{j,k\}}}\right)\right)=D_j
\end{equation}
for all $j\in\hat{n}$,
\begin{equation}\label{two}
\lim_{d\to0_+}\frac{1}{d}\cdot \frac{1}{\sqrt{1+(b_{jk}d)^2}}\cdot
\frac{-1}{2+d\sqrt{1+(b_{jk}d)^2}w_{\{j,k\}}}=S_{jk}
\end{equation}
for all $j\neq k$, $j,k\in\hat{n}$, and finally
\begin{equation}\label{three}
\frac{1}{d(n+du)}=\OO(d)
\end{equation}
as $d\to 0_+$. To fulfil (\ref{two}) one can choose
\begin{equation}\label{par w}
w_{\{j,k\}}(d):= -\frac{1}{S_{jk}}\cdot
\frac{1}{d^2}-\frac{2}{d}\,,
\end{equation}
which makes sense because $S_{jk}\ne 0$ by assumption, since then
the limit equals
 $$
\lim_{d\to0_+}\frac{1}{1+\OO(d^2)}\cdot
\frac{-1}{2d+(1+\OO(d^2))\left(-\frac{1}{S_{jk}}-2d\right)}
=S_{jk}\,.
 $$
With the choice (\ref{par w}) taken into account the condition
(\ref{one}) will be satisfied provided $v_j+\frac{1}{d}(\#N_j+1)
-\sum_{k\in N_j}S_{jk}=D_j$, i.e.
\begin{equation}\label{par v}
v_j(d):=D_j-\frac{\#N_j+1}{d}-\sum_{k\in N_j}S_{jk}\,.
\end{equation}
Finally, the last requirement will be satisfied, e.g., if the
expression equals $d$ which is true if
\begin{equation}\label{par u}
u(d):=\frac{1}{d^3}-\frac{n}{d^2}\,.
\end{equation}
Summarizing the argument we conclude that choosing the parameters
in the described approximation according to (\ref{par
w})--(\ref{par u}) we get in the limit the generic boundary
conditions (\ref{3}). We conjecture that such an approximation
would again converge in the norm-resolvent topology.

\section*{Acknowledgments}

The research was supported by the Czech Academy of Sciences and
Ministry of Education, Youth and Sports within the projects
A100480501 and LC06002.


\end{document}